\documentclass[aps,pre,twocolumn,american,showpacs,nobibnotes,footinbib]{revtex4-1}
\usepackage[T1]{fontenc}
\usepackage[latin9]{inputenc}
\usepackage{color}
\usepackage{amsmath}
\usepackage{amssymb}
\usepackage{graphicx}

\makeatletter
\@ifundefined{textcolor}{}
{%
 \definecolor{BLACK}{gray}{0}
 \definecolor{WHITE}{gray}{1}
 \definecolor{RED}{rgb}{1,0,0}
 \definecolor{GREEN}{rgb}{0,1,0}
 \definecolor{BLUE}{rgb}{0,0,1}
 \definecolor{CYAN}{cmyk}{1,0,0,0}
 \definecolor{MAGENTA}{cmyk}{0,1,0,0}
 \definecolor{YELLOW}{cmyk}{0,0,1,0}
}

\makeatother

\begin{document}

\title{Critical behaviour of the $XY$-rotors model on regular and small
world networks}

\author{Sarah De Nigris}
\email{denigris.sarah@gmail.com}
\author{Xavier Leoncini}

\affiliation{Aix Marseille Universit\'e, CNRS, CPT, UMR 7332, 13288 Marseille, France\\
 Universit\'e de Toulon, CNRS, CPT, UMR 7332, 83957 La Garde, France}
\begin{abstract}
We study the $XY$-rotors model on small networks whose number of
links scales with the system size $N_{links}\sim N^{\gamma}$, where
$1\le\gamma\le2$. We first focus on regular one dimensional rings in the microcanonical ensemble.
For $\gamma<1.5$ the model behaves like short-range one and no phase
transition occurs. For $\gamma>1.5$, the system equilibrium properties
are found to be identical to the mean field, which displays a second
order phase transition at a critical energy density $\varepsilon=E/N,~\varepsilon_{c}=0.75$. Moreover for $\gamma_{c}\simeq1.5$
we find that a non trivial state emerges, characterized by an infinite
susceptibility. We then consider small world networks, using the Watts-Strogatz
mechanism on the regular networks parametrized by $\gamma$. We first
analyze the topology and find that the small world regime appears
for rewiring probabilities which scale as $p_{SW}\propto1/N^{\gamma}$.
Then considering the $XY$-rotors model on these networks, we find
that a second order phase transition occurs at a critical energy $\varepsilon_{c}$
which logarithmically depends on the topological parameters $p$ and
$\gamma$. We also define a critical probability $p_{MF}$, corresponding
to the probability beyond which the mean field is quantitatively recovered,
and we analyze its dependence on $\gamma$.
\end{abstract}


\maketitle

\section{Introduction}

Real-life networks are of finite size, loopy and display heavy correlations.
This complexity represents a challenge from several points of view:
first it is computationally expensive when attempting to investigate
the network topology and to simulate dynamical systems upon it; moreover
it becomes rapidly intractable analytically and one is obliged to
make assumptions in order to simplify the picture and perform calculations.
If this effort is crucial to practically afford problems, it also
embeds a deeper question: facing the network complexity and their
omnipresence in real world, it is fundamental to make the distinction
between the essential variables which are able to catch the topology
main features and those details which are unessential for a minimal
though complete description. One of those very fruitful simplifications
is \emph{sparseness}, i.e. the networks considered have in general
a few links per vertex while the network size tends to infinity. More
precisely, a network is \emph{sparse} if $k/N\rightarrow0$ when $N\rightarrow\infty$,
$k$ being the average degree. This basic hypothesis leads to a crucial
consequence: locally, the network can be approximated by a tree, which
means the absence of\emph{ finite loops, }i.e. finite closed
paths, among the vertices. Sparseness and the local tree-likeness
proved essential to analytical studies of dynamical systems on networks:
we cite, focusing just on small-world networks, studies on the Ising
model \cite{exactsolIsing_lopes2004,herrero2002ising}, percolation
\cite{newman1999scaling} and, more recently, on the Kuramoto model
(for a more complete overview, see \cite{dorogovtsev2008critical}).
Therefore, the advantage in terms of numerical computation is evident:
in general, both numerical studies investigating the network topology
\cite{barrat2000properties,newman2000mean} and critical phenomena
on networks \cite{exactsolIsing_lopes2004,herrero2002ising,kim_smallworld2001,medvedyeva2003dynamic}
exploit the assumption of sparseness in its strongest form, taking
the degree as constant. Nevertheless, increasing the links density,
networks exist which are still sparse, fulfilling the aforementioned
condition but they cannot no longer ensure the tree-likeness because
of the heavy presence of loops. It could be argued hence that the
links density could play a non negligible role both on the topological
properties of those networks and on dynamical models defined upon
them. Indeed it is the case of the $XY$-rotors model on regular one-dimensional
chains: we show that the passage between a sparse network in the sense
of $k=\mathcal{O}(1)$ and a dense one ($k=\mathcal{O}(N)$) implies
the emergence of a new metastable state for which the thermodynamic
order parameter does not relax at equilibrium \cite{deNigris2013}.
The links density hence triggers a non trivial effect on the thermodynamic
behavior of the $XY$ model, which by itself is known for possessing
a rich phenomenology investigated in several numerical studies \cite{jain1986monte,janke1991numerical,kim2007novel,lee1984discrete,loft1987numerical,mccarthy1986numerical}
on 2 and 3-D lattices. In particular, we would like to recall, as
an example among many others, that the two dimensional case with nearest
neighbors coupling is characterized by the famous Berezinskii-Kosterlitz-Thouless
phase transition \cite{kosterlitz2002ordering,leoncini1998hamiltonian},
which implies the correlation function to switch from a power law
decay at low temperatures to an exponential one in the high temperatures
regime. In the mean field limit, the $XY$ model, called the Hamiltonian
Mean Field (HFM) model in this case, displays as well a wide variety
of behaviors this complexity being strongly entangled with the lack
of additivity. Among its peculiarities we cite the presence of a second
order phase transition of the magnetization \cite{Campa2009} and,
even more noteworthy, the presence of non equilibrium quasi-stationary
states of diverging duration in the thermodynamic limit \cite{antoniazzi2007maximum,ettoumi2011linear,latora2002fingerprints,chavanis2005dynamics,Levin2012_ergodicity,levin2011_coreHalo}.
More recently, those models have been challenged to face more complex
network topologies: for instance, studies exist concerning the HMF
model on random graphs \cite{ciani2011long} where, varying the links
density, a second order phase transition of the global magnetization
is recovered for every density value in the thermodynamic limit. Furthermore,
studies of the $XY$ model on small world networks \cite{kim_smallworld2001,medvedyeva2003dynamic}
proved that this lattice topology supports as well complex thermodynamical
responses of the model: a mean field transition of the order parameter
is retrieved and its critical energy seems to depend on the network
parameters. \\
The present work inscribes itself on this line as we will focus, on
first instance, on regular networks and then we will shuffle this
regular topology with the introduction of a controlled amount of randomness.
The first part of the paper being on regular networks, we detail the
analytical calculations presented in \cite{deNigris2013} showing
that tuning the link density allows to pass from a short-range regime
to a long-range one. The analytical approach is preceded in Sec. \ref{sec:Thermodynamic-Behaviour regular}
by the results of numerical simulations which are as well more extensively
illustrated than in \cite{deNigris2013}. Furthermore, we show that
it exists between those two regimes a peculiar metastable state characterized
by huge fluctuations of the order parameter. We then address, in the
second part of the paper, small-world networks using the Watts-Strogatz
model \cite{watts_strogatz1998SW} aiming to shed light on the interplay
between the link density and the injection of randomness in the network.
In his regime we first investigate, acting on the links density $\gamma$
and on the rewiring probability $p$, the crossover from the regular
lattice to the small-network topology. In Sec.\ref{sec:Thermodynamic-Behaviour}
we consider the dynamics of the $XY$-rotors model on small-world
networks and we show how the emergence of global coherence, via a
mean field phase transition of the order parameter, strongly depends
on the topological conditions fixed by $p$ and $\gamma$. Furthermore,
we discuss in the last part how this influence turns out to be \emph{quantitative},
affecting the critical energy $\varepsilon_{c}$ at which the phase transition
occurs.

\section{The $XY$-Rotors Model\label{sec:The-XY-Rotors-Model} }

The $XY$-rotors model describes a set of $N$ spins interacting pairwise:
each spin is fixed on the sites of a one dimensional ring and it is
assigned with two canonically conjugated variables $\{\theta_{i},p_{i}\}$,
$\theta_{i}\in\left[-\pi;\pi\right]$ being a rotation angle. The
$XY$ Hamiltonian reads \cite{antoni1995clustering,leoncini1998hamiltonian}:
\begin{equation}
H=\sum_{i=1}^{N}\frac{p_{i}^{2}}{2}+\frac{J}{2k}\sum_{i,j}^{N}a_{i,j}(1-\cos(\theta_{i}-\theta_{j})),\label{eq:potential HMF}
\end{equation}
\\
where $a_{i,j}$ is the matrix encoding the spins connections:
\begin{equation}
a_{i,j}=\begin{cases}
1\,\, \mbox{if\,i$\neq$ j\ and\ are\, connected} \\
0\,\, \mbox{otherwise}
\end{cases}.\label{eq:adiacency matrix}
\end{equation}
We take $J>0$, so that we are in the ferromagnetic case and in the
following $J=1$ as well as the lattice spacing. Finally the $1/k$ factor
in Eq.~(\ref{eq:potential HMF}) ensures that the energy is an extensive
quantity. $k$ is referred to as the \emph{degree} and, to control
the density of links in the network, we define it as:
\begin{equation}
k=\frac{2^{2-\gamma}(N-1)^{\gamma}}{N}\sim2^{2-\gamma}N^{\gamma-1}.\label{eq:degree}
\end{equation}
Practically, we take the integer part of Eq.~(\ref{eq:degree})
since, once fixed $\gamma$ and $N$, $k$ is in general non integer.
Since we assign $k$ links per spin and we set periodic boundary conditions, the system is translationally invariant.
The dynamics are given by the set of Hamilton equations:
\begin{eqnarray}
\dot{\theta_{i}} & = & \frac{\partial H}{\partial p_{i}}=p_{i},\label{eq:eq dynamics}\\
\dot{p_{i}} & = & -\frac{\partial H}{\partial\theta_{i}}=-\frac{J}{k}\sum_{j\in V_{i}}^{N}\sin\left(\theta_{i}-\theta_{j}\right)\nonumber 
\end{eqnarray}
where $V_{i}$, represents the neighbors of rotor $i$. A global parameter,
the magnetization is defined by 
\begin{eqnarray}
\mathbf{M} & =\frac{1}{N} & \left(\begin{array}{cc}
 & \sum\cos\theta_{i}\\
 & \sum\sin\theta_{i}
\end{array}\right)=M\left(\begin{array}{cc}
\cos\varphi\\
\sin\varphi
\end{array}\right)\label{eq:Magnetisation_def}
\end{eqnarray}
in order to have an insight on the macroscopic behavior: we expect
finite values of $M$ to indicate the emergence of a coherent inhomogeneous state 
%\footnote{The term ``inhomogeneous`` refers to the one particle PDF which is, in the ordered phase, inhomogeneous in the $\theta$ space, since it exists a preferred direction for the magnetization.} 
, while a vanishing magnetization signals the absence of long-range
order. We first study the response of the total equilibrium magnetization
$M$ to the change of the underlying network via the $\gamma$ parameter.
Practically, for each $\gamma$, we perform simulations within the
microcanonical ensemble, by direct numerical integration of Eqs.~(\ref{eq:eq dynamics})
with the fifth order optimal symplectic integrator described in \cite{mclachlan1999accuracy}.
The initial conditions of angles and momenta are picked from a Gaussian
distributions with identical variance (which corresponds to a low
temperature setting) and, to check the numerical integration, we monitor
the conservation of the two constants of motion preserved by the dynamics:
the energy $E=H$ and the total angular momentum $P=\sum_{i}p_{i}$,
which we have set without loss of generality to $P=0$. Finally the
time step is $\Delta t=0.05$ and we average the thermodynamic quantities
over time only when the system has reached the equilibrium.

\section{Thermodynamic Behavior on Regular Lattices\label{sec:Thermodynamic-Behaviour regular}}

\subsection{: Numerical Computation\label{sub::-Numerical-Computation}}

The regular network that we take into account is a one-dimensional
chain of $N$ spins (rotors) with periodic boundary conditions for
which each spin is connected to its $k$ nearest neighbors. By tuning
the parameter $\gamma$, $1<\gamma\leq2$ we act on the links density
of the network. For $\gamma=1\,\,(k=2)$ the spins are connected to
their nearest neighbors, while for $\gamma=2\,\,(k=N-1)$ the network
is fully coupled. Heuristically, changing the value of $\gamma$ corresponds
to change the \emph{range} of interaction of each spin. Then two limit
behaviors naturally emerge from this approach: the first is $\gamma\rightarrow$1
in which we expect the system to behave progressively like a one dimensional
short range system with the existence of a continuous symmetry group,
and so without any phase transition . On the other side, the $\gamma\rightarrow2$
limit leads to the mean field regime and we expect the HMF transition
of the magnetization to appear above a specific threshold of degree.
We find this boundary value for $\gamma=1.5$ so that the two aforementioned
limits translate more precisely in two intervals $\gamma<1.5$ and
$\gamma>1.5$. Practically, for each $\gamma$ value, we monitor the
average magnetization $\overline{M(N,\varepsilon)}$ (the bar indicates
the temporal mean) for different sizes $N$ and for every energy density
$\varepsilon=E/N$ in the physical range. The temporal mean is computed
on the second half on the simulations: we start with the Gaussian
initial conditions described in Sec. \ref{sec:The-XY-Rotors-Model}
and we simulate the dynamics, calculating the magnetization at each
time step. When the system reaches a stationary state for the magnetization,
we take the temporal mean as the equilibrium value.

We start our analysis with the $\gamma<1.5$ interval. The simulations
are displayed 
\begin{figure}
(a)\includegraphics[width=7.5cm, keepaspectratio]{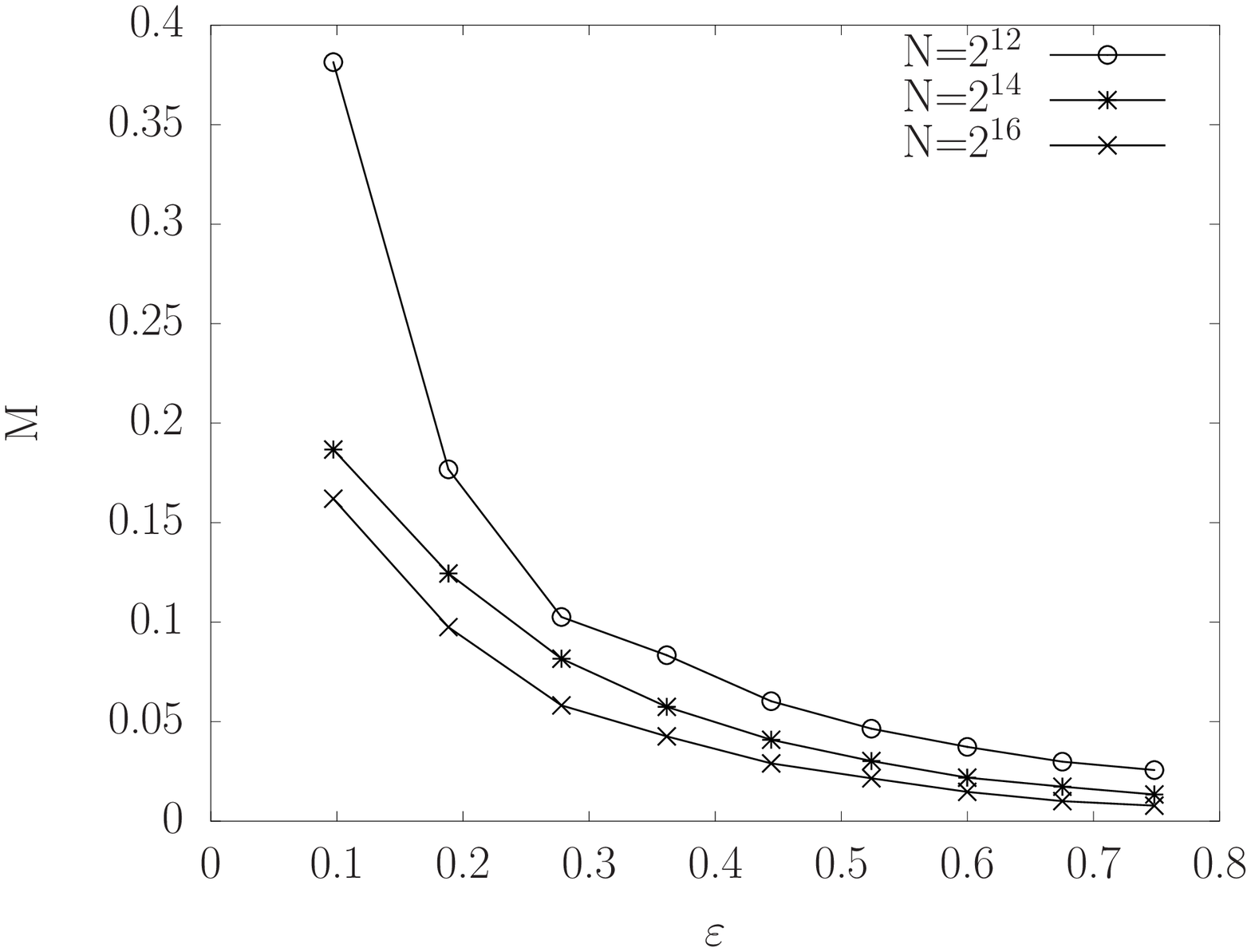}\\
(b)\includegraphics[width=7.5cm, keepaspectratio]{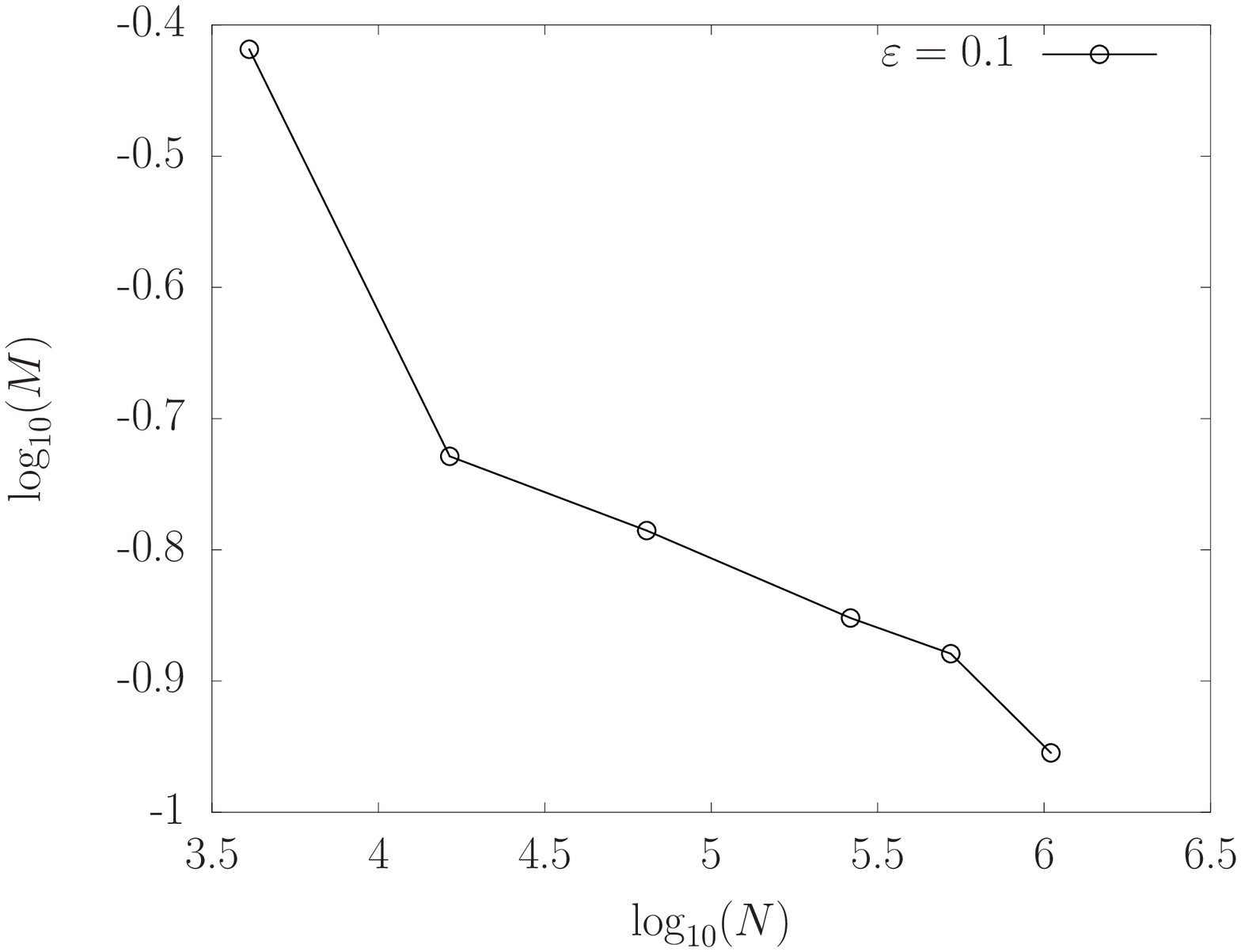}\caption{\label{fig:low gamma}(a) Equilibrium magnetization versus energy
density for $\gamma=1.25$ and different sizes. The error bars are
of the size of the dots; (b) Residual magnetization for $\gamma=1.25$
at $\varepsilon=0.1$ versus the system size.}
\end{figure}
 in Figs.~\ref{fig:low gamma}a-b and as mentioned the magnetization
smoothly vanishes with the energy (Fig. \ref{fig:low gamma}a).
We have to recall that for low energies, the magnetization can
be non-zero as a finite size effect, so the results displayed should
depend on the system size. This is confirmed in Fig.~\ref{fig:low gamma},
where the trend for the magnetization to vanish with increasing the
size is exhibited. To check with even larger sizes, we consider in
Fig.~\ref{fig:low gamma}b the magnetization for a small energy density
$\varepsilon=0.1$ and several sizes. The results clearly point out
that the magnetization vanishes in the thermodynamic limit. When looking
at relaxation scales, we found that larger sizes took more time to
relax to equilibrium. So typically in our simulations we take as final
time $t_{f}=20000$ for sizes up to $N=2^{16}$ and for $N>2^{18}$~~$t_{f}=30000$. 

\begin{figure}
\includegraphics[width=8cm, keepaspectratio]{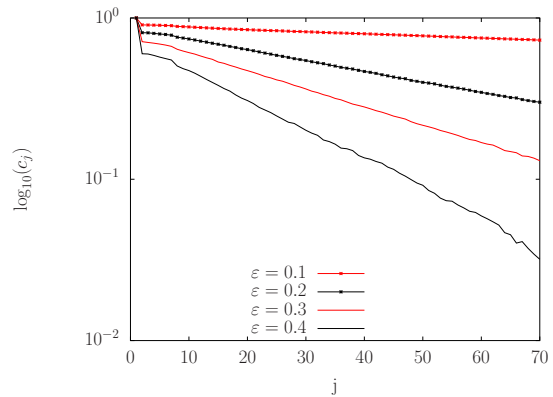}\caption{(color online) Correlation function $c_{j}$ for $\gamma=1.25$ and
$N=2^{14}$.\label{fig:correlations 1.25}}

\end{figure}

Given these numerical results, we conclude that in the $\gamma<1.5$
interval, the system is short-ranged and the Mermin-Wagner theorem
applies imposing the order parameter to vanish. Nevertheless, if long-range
order is not possible, quasi long-range could still entail an infinite
order phase transition of the correlation function, like in the two
dimensional $XY$ model with nearest neighbors interactions. We recall
that this particular type of critical phenomenon, first detected by
Berezinskii, Kosterlitz and Thouless \cite{kosterlitz2002ordering},
is characterized by two different types of decay of the correlation
function with distance: a power law or an exponential decay, respectively
for low and high temperatures. In order to look for such possibility
we computed the correlation function 
\[
c(j)=\frac{1}{N}\sum_{i=1}^{N}\cos(\theta_{i}-\theta_{i+j\left[N\right]})\:,
\]
for every $\varepsilon$ in the considered range. The results shown
in Fig. \ref{fig:correlations 1.25}
 indicate that the decay behavior is also exponential for low energies,
demonstrating the absence of the aforementioned phase transition.
This could have been anticipated from the fact that finite size effects
on the magnetization even though present were small, but possible
tricky effects of the boundary conditions could come into play, so
it was worthwhile checking.

To summarize our result, we can conclude that, for $\gamma<1.5$,
the spin degree is still too low for the system to show long-range
or quasi long-range. Interestingly, the short range behavior is still
at play even for configurations like $\gamma=1.4$ where each spin
is under the influence of quite an important neighborhood since $k\propto N^{0.4}$
in this case.

\begin{figure}
\includegraphics[width=8cm, keepaspectratio]{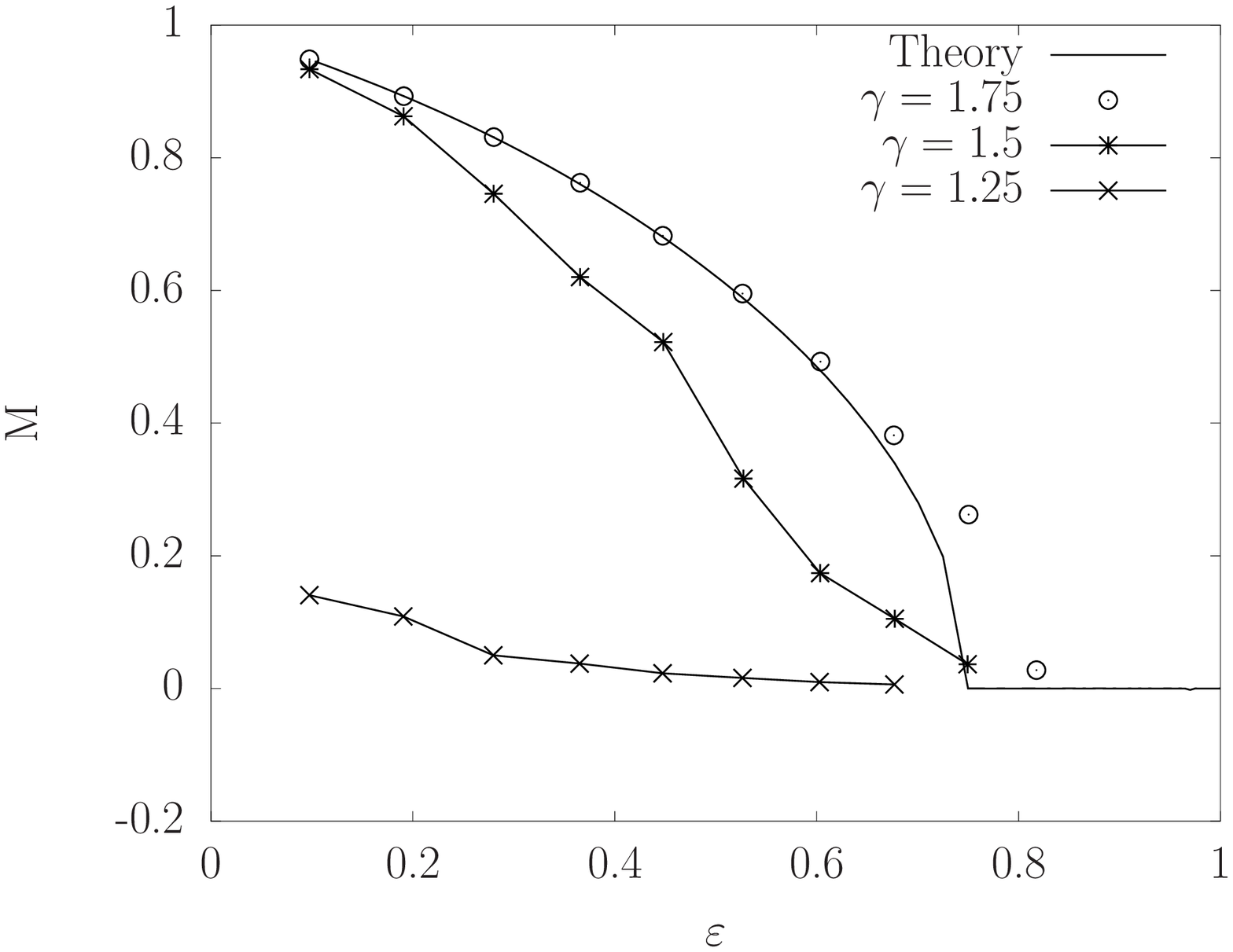}\caption{Equilibrium magnetization for $N=2^{16}$ and different $\gamma$.
For $\gamma\neq1.5$ the error bars are of the size of the dots. \label{fig:various gamma}}

\end{figure}

\begin{figure}
(a)\includegraphics[width=8cm, keepaspectratio]{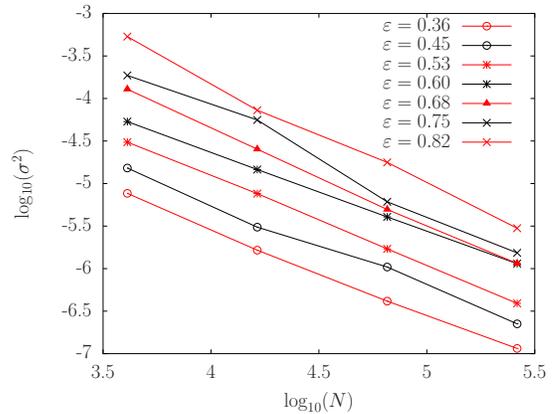}\\

(b)\includegraphics[width=8cm, keepaspectratio]{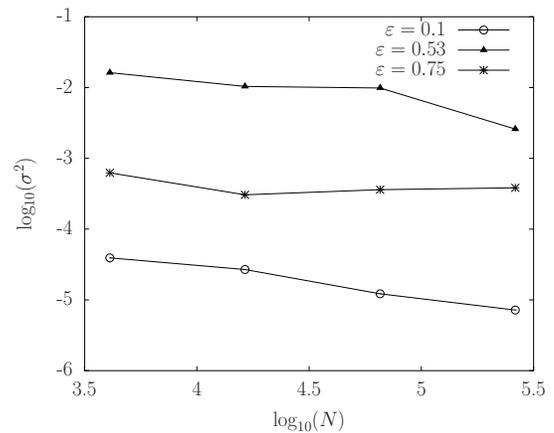}\caption{(color online) Scaling of the magnetization variance $\left\langle \sigma^{2}\right\rangle $
with the size for $\gamma=1.75$ (a) and $\gamma=1.5$ (b).\label{fig:variance 1.5}}

\end{figure}

Taking now in account the symmetric interval $\gamma>1.5$, the spins
are connected enough to allow a coherent state to emerge: in Fig. \ref{fig:various gamma}
the magnetization undergoes a second order phase transition at $\varepsilon_{c}=0.75$
which is well described by the HMF analytical curve. Again, around
the delicate zone of the phase transition, finite size effects induce
a shift between the theoretical prediction of the HMF and the simulations,
but they can be smoothed down increasing the size. We recall that
this phenomenon is also present for the full coupling $\gamma=2$.
As a consequence we then find that even with a degree remarkably inferior
(e.g. for $\gamma=1.6$ ) than the full coupling condition, each spin
possesses enough connections to trigger the global behavior of the
system and give a finite magnetization (at low energies). Of course,
in both the intervals $\gamma\lessgtr1.5$, the equilibrium magnetization
is still affected by fluctuations because of the finite size effects.
To monitor these we measured the magnetization variance $\sigma^{2}=\overline{(M-\overline{M})^{2}}$
and we show in Fig.~\ref{fig:variance 1.5} that it scales with the
system size like 
\begin{equation}
\sigma^{2}\propto1/N.\label{eq:scaling sigma}
\end{equation}
This scaling is the one expected for the equilibrium state thus confirming
that the values in Figs.~\ref{fig:low gamma}a- \ref{fig:various gamma}
are representative of such state.

\begin{figure}
(a)\includegraphics[width=8cm, keepaspectratio]{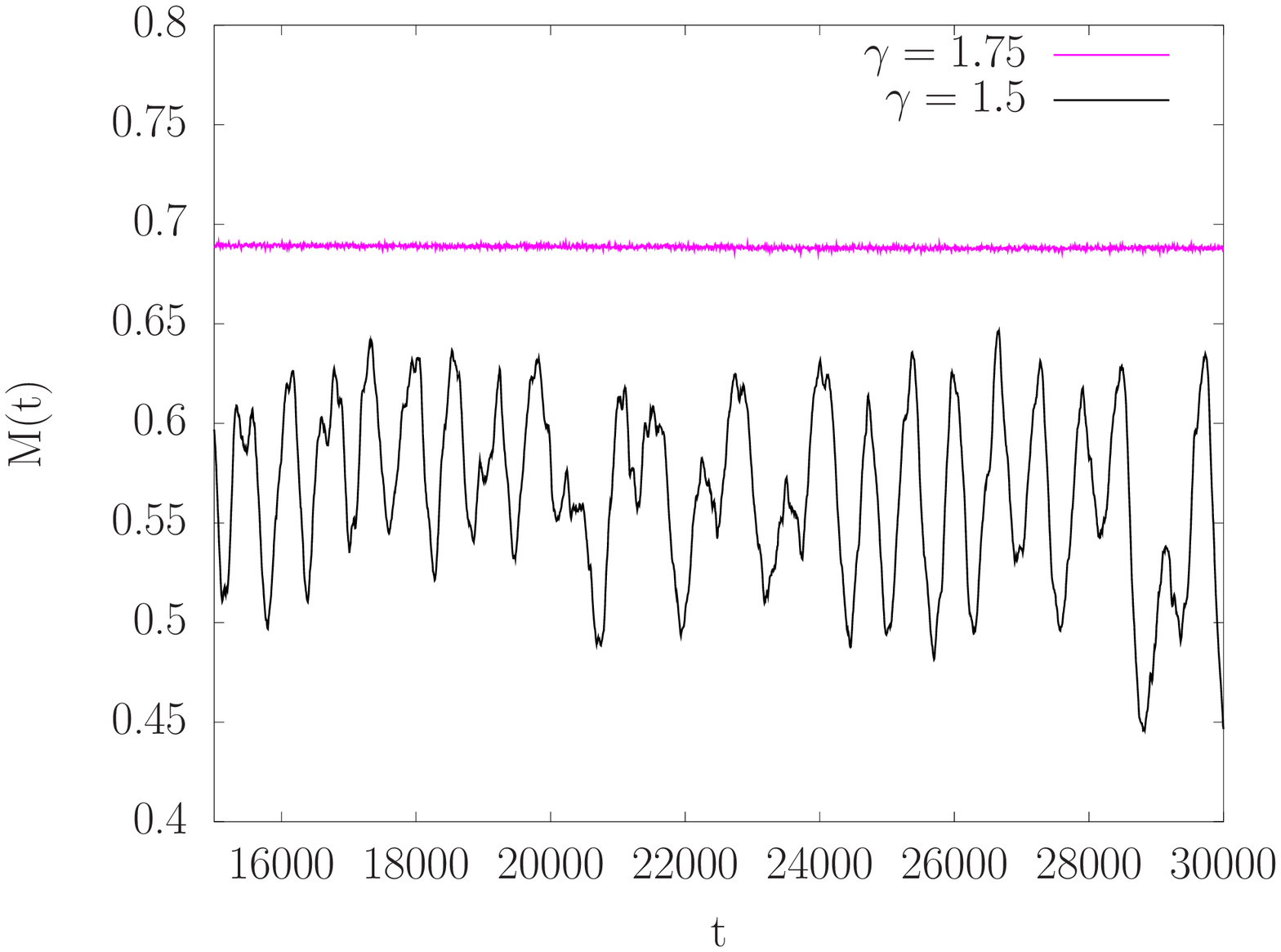}\\
(b)\includegraphics[width=8cm, keepaspectratio]{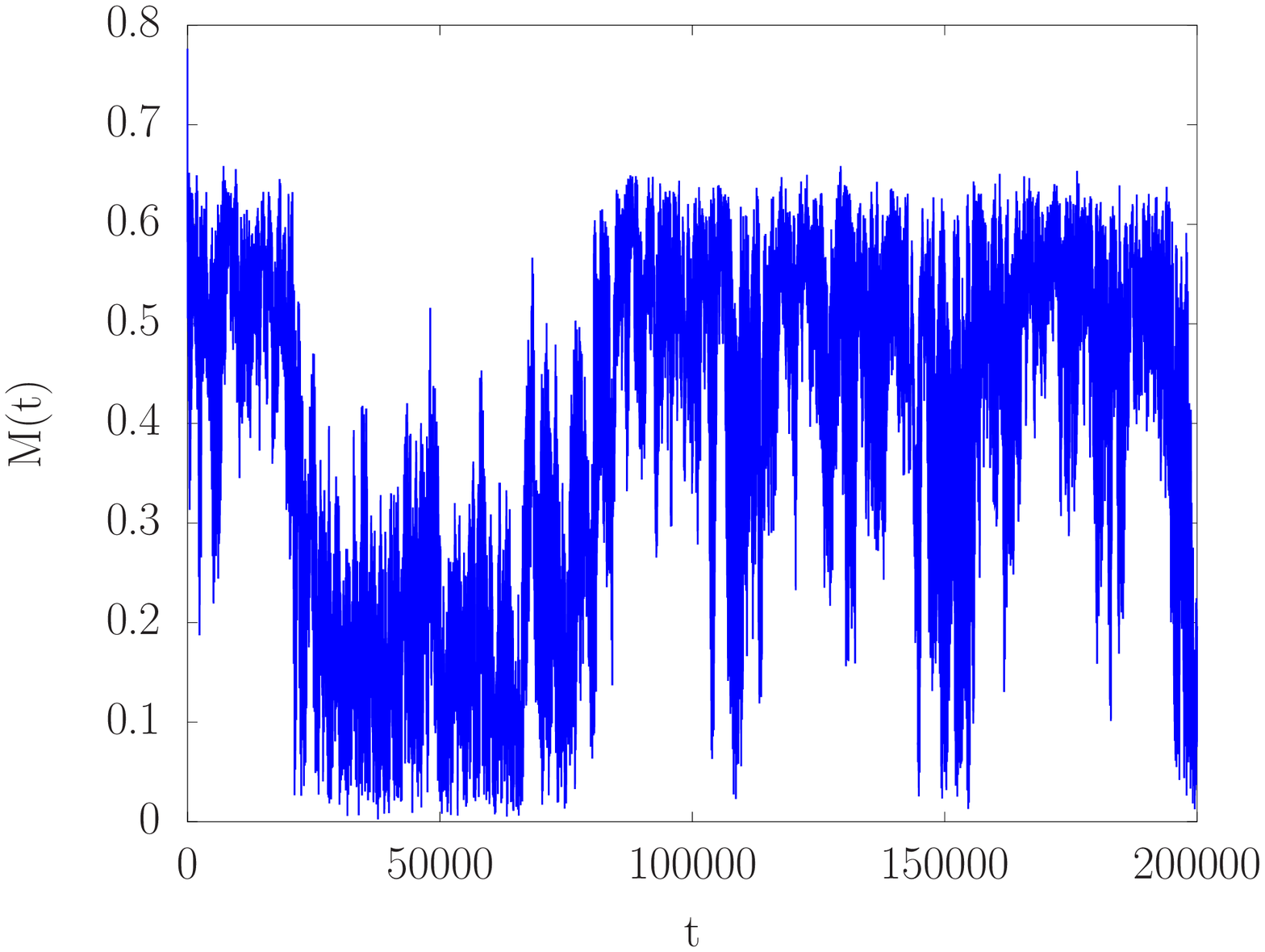}\\
(c)\includegraphics[width=8cm, keepaspectratio]{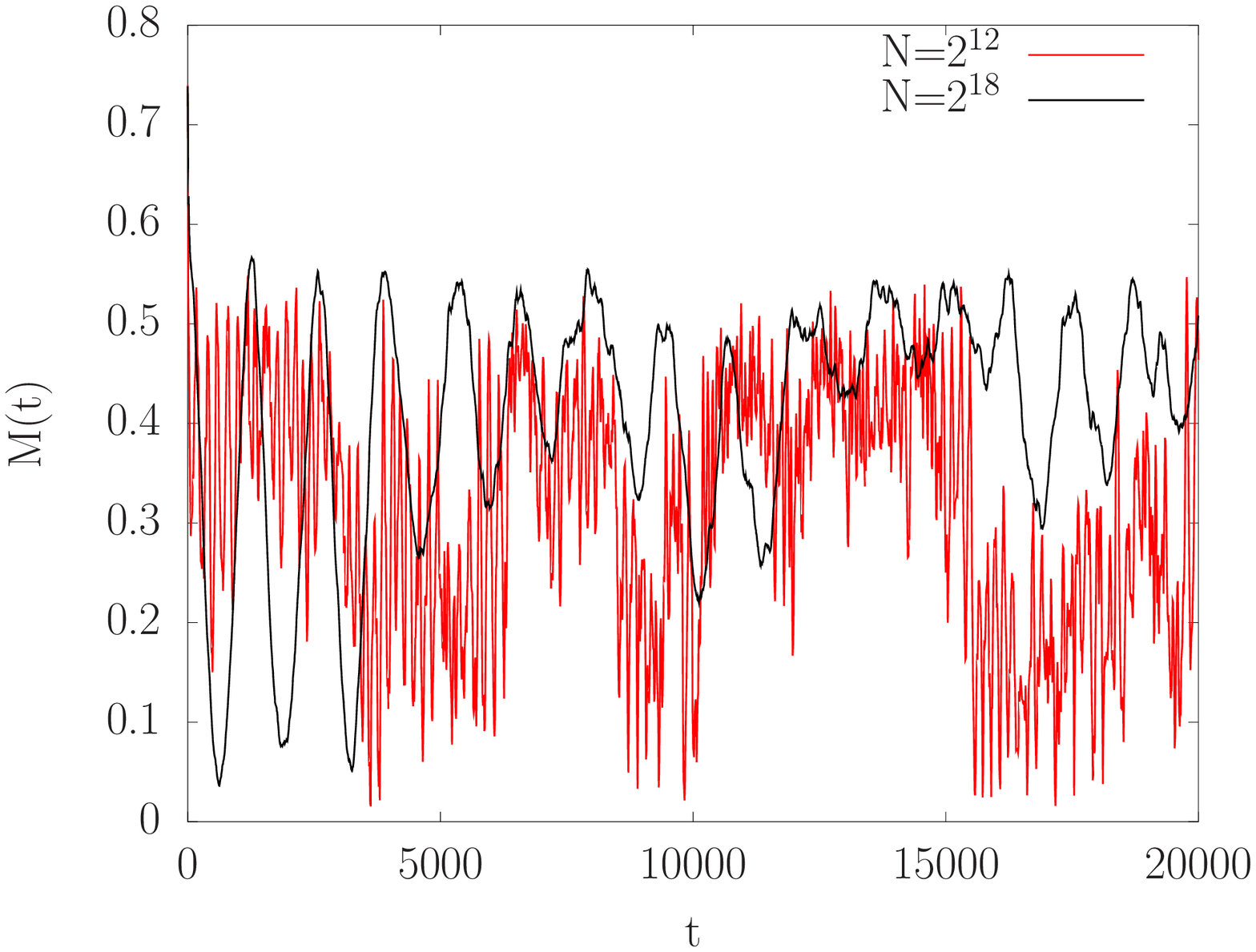}\\
(d)\includegraphics[width=8cm, keepaspectratio]{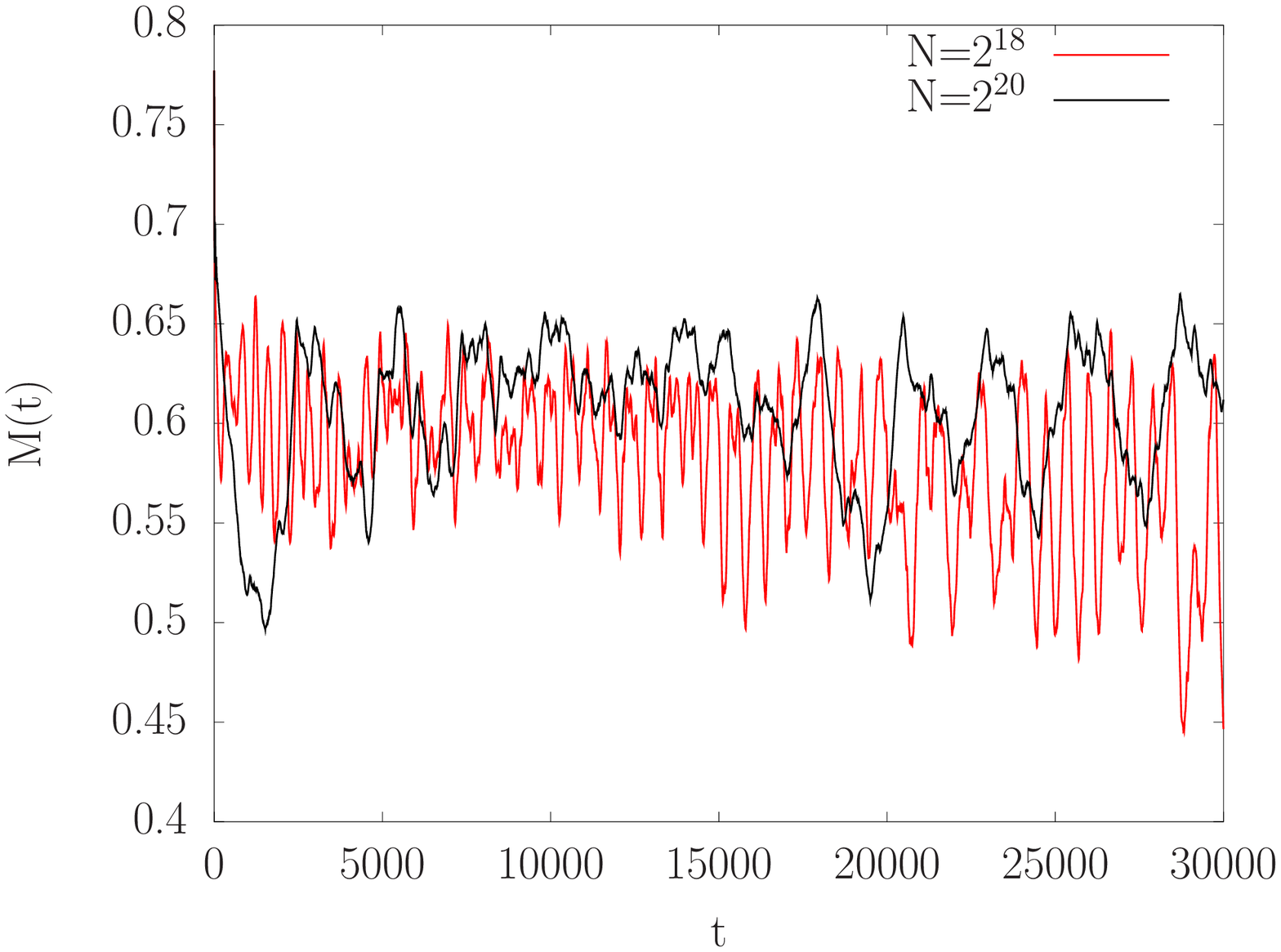}

\caption{(color online) Time series for the magnetization with (a) $N=2^{18}$, $\varepsilon=0.60$; (b) $N=2^{12}$; $\varepsilon=0.44$; $T_{f}=200000$; 
(c) Comparison of the fluctuations amplitude of $N=2^{12}$ and $N=2^{18}$ with $\varepsilon=0.52$ and (d) of $N=2^{18}$ and $N=2^{20}$ 
with $\varepsilon=0.44$.\label{fig:fluctuations}}

\end{figure}

Given the results presented in the previous discussions, a natural
critical value appears, which characterizes the shift from the short
range picture to the long range one: $\gamma_{c}\simeq 1.5$. We decided
to investigate the system behavior at this critical threshold imposing
$\gamma=\gamma_{c}$. In fine, we expect that the system will be in
a peculiar state by itself which cannot be labeled as short or long
ranged. Results are depicted in Fig.~\ref{fig:various gamma}. We
observe that for low energies, $0.3\lesssim\varepsilon\leq0.75$,
the averaged magnetization is finite even when increasing the size
but it remains lower than the mean field value. The effect is clearer
when we look at its temporal behavior. It is indeed totally different
than in the other two regimes and the order parameter $M$ shows large
fluctuations which are orders of magnitude larger than for the other $\gamma$
regimes. We show in Fig.~\ref{fig:fluctuations} a comparison a time
series for the same energy and system size and different values of
$\gamma$, namely $\gamma=1.75$ which displays a finite magnetization
with small fluctuations and the one $\gamma=1.5$, with large fluctuations.
In order to control the fact that these fluctuations are not an artifact
of our initial conditions and that it is likely that the system does
not relax on larger timescales than the previous configurations, we
considered computation times up to a final time $t_{f}=200000$. Results
are presented in Fig.~\ref{fig:fluctuations}, where it appears that
this regime with large fluctuations persists. We recall that for $\gamma\lessgtr1.5$
the simulation time was at most $t_{f}=30000$ and it was enough to
reach a stationary state. Proceeding further, we notice that the amplitude
of these fluctuations is not dependent on system size. We compare
for instance $N=2^{12}$ to $N=2^{18}$ in Fig.~\ref{fig:fluctuations}
c and we conclude that for the aforementioned energies there is no
significant amplitude decrease with system size. More precisely, 
if we consider the variance $\sigma^{2}$ as before, it appears that
the scaling of the variance mentioned in Eq.~(\ref{eq:scaling sigma})
and coherent with in the $\gamma\ne1.5$ regimes, is substituted by
a flat behavior increasing $N$ (see the results in Fig.~\ref{fig:variance 1.5}b).
It is worth noticing that the influence the system size can be retrieved
not in the fluctuations amplitude but in the typical fluctuation time
scale. In Figs. \ref{fig:fluctuations}c-d, it becomes obvious that
fluctuations appear to slow down with the system size. This time-scale
dependence on the size is reminiscent of out of equilibrium behavior
in systems with long range interactions, namely the lifetime of the
Quasi Stationary States (QSS) \cite{chavanis2008out,van2010stationary,ettoumi2011linear,Ettoumi2013}
and further investigations are ongoing to shed light on this effect
and on its potential analogy with the HMF results.

Heuristically, for $\gamma=\gamma_{c}$ it is like the if each spin
does not possess enough connections to create a global order and establish
the mean field but, nevertheless, the degree is sufficiently high
($\gamma=1.5$ corresponds to $k=\sqrt{N}$) to avoid the vanishing
of the order parameter in the thermodynamic limit. The resulting behavior
is reminiscent of a bistable regime oscillating between the $M=0$
configuration and the mean field value, which corresponds to a finite
magnetization, we thus may expect some kind of intermittent behavior.
In fine, the flatness of the variance suggests moreover that we observe
a state with infinite susceptibility $\chi$ considering its canonical definition
\begin{equation}
\chi\sim\lim_{N\rightarrow\infty}N\sigma^{2}.\label{eq:chi}
\end{equation}
To conclude our analysis in symmetry with the $\gamma\lessgtr1.5$
cases, we looked for a signature of this non trivial state in the
correlation function but the fluctuations heavily affect it too so
that it oscillates without showing a proper scaling.

\subsection{Analytical Calculation\label{sub:Analytical-Calculation}}

\begin{figure}
\includegraphics[width=8cm, keepaspectratio]{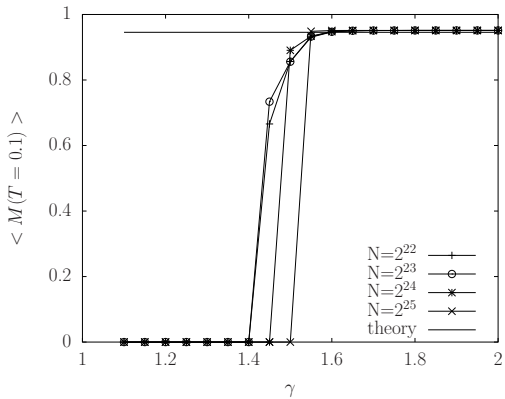}\caption{Analytical magnetization $\left\langle M\right\rangle $
from Eq.~(\ref{eq:final magnetisation}) for $T=0.1$ versus $\gamma$.
Theory refers to the exact analytical solution of the HMF model.\label{fig:magn analitical}}
\end{figure}

The numerical investigations illustrated point out that the degree
triggers the shift from the pure one dimensional topology to the mean
field frame. We now tackle this issue analytically aiming to retrieve
the influence of the topology, encoded in the adjacency matrix $a_{i,j}$
(Eq.~(\ref{eq:adiacency matrix})), in the thermodynamic properties.
We thus compute the magnetization in the low energy regime and check
if the correct behavior is recovered, namely a zero magnetization
for $\gamma<1.5$ and a finite value for $\gamma>1.5$. At low energies
we have a clear separation between the magnetization values, with
$M=0$ and the mean field one, in which as $\varepsilon\rightarrow0\,\, M\rightarrow1$.
In this limit, due to the ferromagnetic coupling it is natural to
assume the differences $\theta_{i}-\theta_{j}$ are small when $a_{i,j}=1$
so that the connected spins are mostly aligned in order to minimize
the free energy. We can hence develop the Hamiltonian at the leading
order: 
\begin{equation}
H=\sum_{i}\frac{p_{i}^{2}}{2}+\frac{J}{4k}\sum_{i,j}a_{i,j}(\theta_{i}-\theta_{j})^{2}\:,\label{eq:Hamilton_dl}
\end{equation}
so that our system reduces to a collection of oscillators connected
by $a_{i,j}$. We then choose to represent the spin field as
a superposition of modes, following the recipe given in refs \cite{leoncini1998hamiltonian,leoncini2001dynamical}:
\begin{equation}
\begin{array}{c}
\theta_{i}=\sum_{l=0}^{N-1}\alpha_{l}(t)\cos(\frac{2\pi li}{N}+\phi_{l})\\
p_{i}=\sum_{l=0}^{N-1}\dot{\alpha_{l}}(t)\cos(\frac{2\pi li}{N}+\phi_{l})
\end{array}.\label{eq:representation}
\end{equation}
In Eq.~(\ref{eq:representation}), we sum over $N$ modes so that
the change of variables is linear and we observe that, given the periodic
boundary conditions, it just corresponds to perform a discrete Fourier
transform. The amplitudes $\alpha_{l}$ are, in our approach, the
information carriers of the temporal behavior, hence the representation
of the momenta $p_{i}$ is related to the one of the angles via the
first Hamilton equation $p_{i}=\dot{\theta_{i}}$. The phases $\phi_{l}$
are randomly distributed on the circle to ensure that the momenta
$p_{i}$ are Gaussian distributed in the limit $N\rightarrow\infty$
as theoretically predicted for the microcanonical ensemble. Following
the approach described in \cite{leoncini2001dynamical}, if we consider
different sets of phases $\{\phi_{l}\}_{m}$ labeled by $m$ we can
interpret each set as a realization of the system, i.e. a trajectory
in the phase space. Hence the process of averaging on random phases
would correspond to ensemble averaging and leads to \emph{dynamic}
equations which, nevertheless, embed information about the \emph{thermodynamic
state} of the system, via the phase averaging. If we now inject Eq.~(\ref{eq:representation}) in the Hamiltonian (\ref{eq:Hamilton_dl})
we obtain for the kinetic part $K$:
\begin{equation}
\frac{\left\langle K\right\rangle }{N}=\frac{1}{N}\left\langle \sum_{i}\frac{p_{i}^{2}}{2}\right\rangle =\frac{1}{4}\sum_{l}\dot{\alpha}_{l}^{2},\label{eq:average kinetic}
\end{equation}
where $\left\langle ...\right\rangle $ stands for the average over
random phases. In Eq.~(\ref{eq:average kinetic}) we used the relation:
\[
\left\langle \cos(k_{i}+\phi_{i})cos(k_{j}+\phi_{j})\right\rangle =\frac{1}{2}\delta_{i,j}.
\]
 For the potential, we have that the adjacency matrix $a_{i,j}$
is a circulant one because of the definition of the regular network
given in Sec.~\ref{sec:The-XY-Rotors-Model} which is translationally invariant. Hence we can diagonalize
it, obtaining a real spectrum $\{\lambda_{j}\}$ since $a_{i,j}$ is a real symmetric matrix. The spectrum analytical expression in general reads:
\begin{equation}
\lambda_{j}=\frac{1}{k}\sum_{l=1}^{N-1}c_{l}e^{\frac{2\pi ijl}{N}},\label{eq:general spectrum}
\end{equation}
where the vector $c_{l}$ is the coefficient vector whose permutations compose the matrix $a_{i,j}$. 
Because of the two symmetries $c_{l}=c_{N-l}$ and $e^{\frac{2\pi ij(N-l)}{N}}=e^{\frac{-2\pi ilj}{N}}$, Eq.~(\ref{eq:general spectrum}) can be splitted in two sums:
\begin{equation}
 \lambda_{j}=\frac{1}{k}\left(\sum_{l=1}^{\frac{N}{2}}c_{l}e^{\frac{2\pi ijl}{N}}+\sum_{l=1}^{\frac{N}{2}}c_{l}e^{\frac{-2\pi ijl}{N}}\right),
\end{equation}
which can hence be written as the sum of the real parts:
 
\begin{equation}
\lambda_{j}=\frac{2}{k}\sum_{l=1}^{k/2}\cos(\frac{2\pi lj}{N})=\frac{1}{k}\left[\frac{\sin[(k+1)j\pi/N]}{\sin(j\pi/N)}-1\right],\label{eq:spectrum}
\end{equation}
where $k$ is the spin degree of Eq.~(\ref{eq:degree}). To the leading
order the potential will hence take the form:
\begin{equation}
\frac{V}{N}=\frac{1}{4kN}\sum_{i,j}a_{i,j}(\theta_{i}-\theta_{j})^{2}=\frac{1}{2}\sum_{l}(1-\lambda_{l})\left|\hat{\theta_{l}}\right|^{2}\label{eq:linearised potential}
\end{equation}
In Eq.~(\ref{eq:linearised potential}) we used the identity 

\begin{equation}
 \frac{1}{kN}\sum_{i,j}a_{i,j}\theta_{i}\theta_{j}=\frac{1}{kN}\Theta^{T}P^{*}DP\Theta=\sum_{l}\lambda_{l}\left|\hat{\theta_{l}}\right|^{2}\label{eq:identity},
\end{equation}
where $\Theta=\left( \theta_{1}...\theta_{N}\right)$ and $a_{i,j}=P^{*}DP$. In the latter equation $D$ is the diagonal form of the adiacency matrix
and $P^{-1}=P^{*}$ since $P$ is unitary. The identity in Eq.~(\ref{eq:identity}) comes 
from the fact that the eigenvectors of a circulant matrix of size $N$
are the columns of the unitary discrete Fourier transform matrix of
the same size. We can then inject in Eq.~(\ref{eq:linearised potential})
the linear waves representation and average over the phases as we
did for the kinetic part of the Hamiltonian:
\[
\frac{\left\langle V\right\rangle }{N}=\left\langle \frac{1}{2}\sum_{l}(1-\lambda_{l})\left|\hat{\theta_{l}}\right|^{2}\right\rangle =\frac{1}{4}\sum_{l}(1-\lambda_{l})\alpha_{l}^{2}.
\]
Having obtained the averaged Hamiltonian $\left\langle H\right\rangle =\left\langle K\right\rangle +\left\langle V\right\rangle $,
we can deduce the \emph{averaged equation of motion,} as anticipated,
via the second Hamilton equation 
\[
\frac{d}{dt}\left(\frac{\partial\left\langle H\right\rangle }{\partial\dot{\alpha_{l}}}\right)=-\frac{\partial\left\langle H\right\rangle }{\partial\alpha_{l}}\:,
\]
and obtain
\begin{equation}
\ddot{\alpha_{l}}=-(1-\lambda_{l})\alpha_{l}=-\omega_{l}^{2}\alpha_{l}.\label{eq:dispersion_relation}
\end{equation}
We have hence an equation for an harmonic oscillator whose frequency
depends on the adjacency matrix spectrum and, consequently, on the
spin degree. We note that this approach is dependent from our low
temperatures approximation, but as mentioned we shall make use of
this since, depending on the value of $\gamma$, we expect
two clearly defined regimes of zero or finite magnetization. Our system
is now completely encoded in terms of wave amplitudes $\{\alpha_{l}\}$
and frequencies $\{\omega_{l}\}$ which can be linked observing that,
at equilibrium, we have the equipartition of the modes ($p_{i}$'s
are Gaussian):
\[
T=\frac{1}{N}\sum_{i}\left\langle p_{i}^{2}\right\rangle =\frac{1}{2}\sum_{l}\alpha_{l}^{2}\omega_{l}^{2}\Rightarrow\alpha_{l}^{2}=\frac{2T}{N(1-\lambda_{l})}.
\]
In order to compute $M$, we apply the same procedure, meaning that
we average over the phases its expression given by Eq.~(\ref{eq:Magnetisation_def})
after having substituted the representation Eq.~(\ref{eq:representation}).
We obtain \cite{leoncini1998hamiltonian}:
\begin{equation}
\left\langle \mathbf{M}\right\rangle =\prod_{l}J_{0}(\alpha_{l})(\cos\theta_{0},\sin\theta_{0}),\label{eq:average magnetisation}
\end{equation}
where $J_{0}$ is the zeroth order Bessel function and $\theta_{0}$
is the average of the angles $\{\theta_{i}\}$, $\theta_{0}=\frac{1}{N}\sum_{i}{\theta_{i}}$. This quantity is conserved
because of the translational invariance, giving a constant total momentum
$P$ which is set at $P=0$ by our choice of initial conditions. As
the final step to evaluate Eq.~(\ref{eq:average magnetisation}),
we recall that we are dealing with a low temperatures approximation
so we can consider that the amplitudes $\alpha_{l}^{2}$ to be small
at equilibrium and in the large system size limit \cite{leoncini2001dynamical}.
This consideration allows to develop at leading order the product
of the Bessel functions and, taking the logarithm of Eq.~(\ref{eq:average magnetisation}),
we finally obtain:
\begin{equation}
\ln\left(\left\langle M\right\rangle \right)=-\sum_{l}\frac{\alpha_{l}^{2}}{4}=-\frac{T}{2N}\sum_{l}\frac{1}{1-\lambda_{l}}.\label{eq:final magnetisation}
\end{equation}
Eq. (\ref{eq:final magnetisation}) conjugates the thermodynamic information
and the topological one because of the matrix spectrum. If, from one side, it actually realizes our purpose of matching these 
two levels of description, now the spectrum in Eq.~(\ref{eq:spectrum}) carries the system complexity, requiring Eq.~(\ref{eq:final magnetisation})
to be evaluated numerically. In Fig. \ref{fig:magn analitical}
 we show, increasing the size, how this approximated expression grasps
the correct asymptotic behavior, giving the mean field value in the
high $\gamma$ regime and vanishing for low $\gamma$. The transition
becomes sharper at $\gamma_{c}\simeq 1.5$ by increasing the size and gives
hence confirmation of its critical signification as already pointed
out by our numerical simulations of Sec.\ref{sec:Thermodynamic-Behaviour regular}.

\section{The Small World Network Model\label{sec:The-Model}}

In Secs.~\ref{sub::-Numerical-Computation}-\ref{sub:Analytical-Calculation}
we considered a regular chain as network topology and we illustrated
how the degree drives the thermodynamic response of the $XY$-model
on those lattices from the short-range regime to the long-range one.
The natural following step to reorganize the topology is now to break
the translational invariance of the regular chain previously considered
and to introduce some \emph{randomness} in how the spins are connected.
In this purpose, we used the Watts-Strogatz model (W-S)\cite{watts_strogatz1998SW}
for small-world networks, which interpolates between a regular network
and a random one by the progressive introduction of random long-range
connections. Following the algorithm devised in \cite{watts_strogatz1998SW},
each link is reconnected with probability $p$ to a randomly chosen
other vertex or is left untouched with probability $1-p$: long-range
connections are hence introduced and the rewiring procedure injects
disorder in the network since $k,$ fixed by Eq.~(\ref{eq:degree})
at the beginning, is non-uniform afterwards. The degree distribution
decays exponentially since the rewiring is performed independently
for every vertex \cite{barrat2000properties}. Moreover, since $k_{i}\approx\left\langle k\right\rangle $,
a W-S network is not locally equivalent, even in the limit case of
$p=1$, to a random graph were eventually isolated vertex exist and
the network is fragmented in many parts \cite{barrat2000properties}.
It is noteworthy for the following to add that the rewiring injects
mainly shortcuts whose length is of the order of the network size $\mathcal{O\left(N\right)},$
so that a fine tuning of the interaction range by the means of the
randomness $p$ is not possible.

\subsection{Network analysis\label{sub:Network-analysis}}

\begin{figure}
(a)\includegraphics[width=8cm, keepaspectratio]{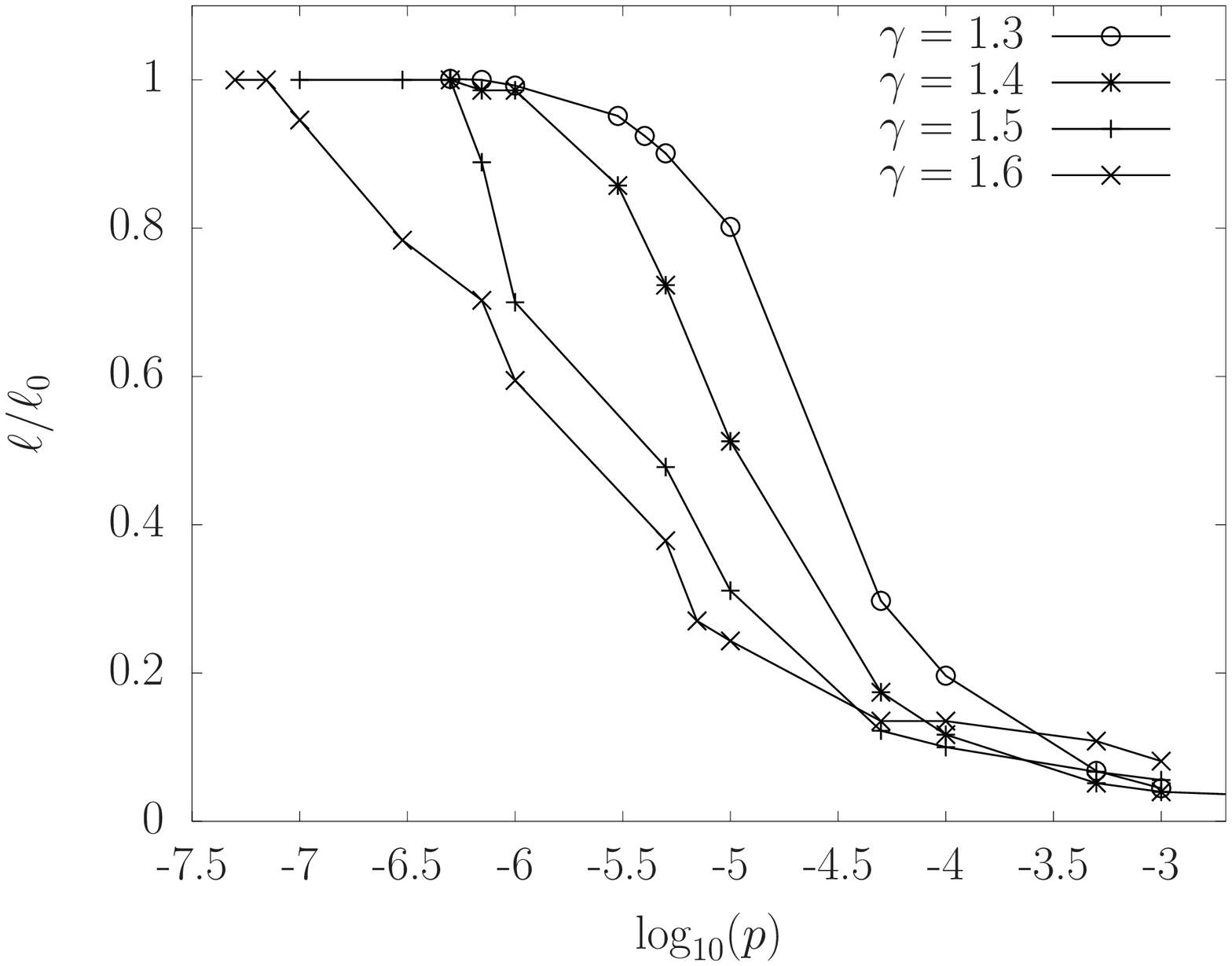}\\
(b)\includegraphics[width=8cm, keepaspectratio]{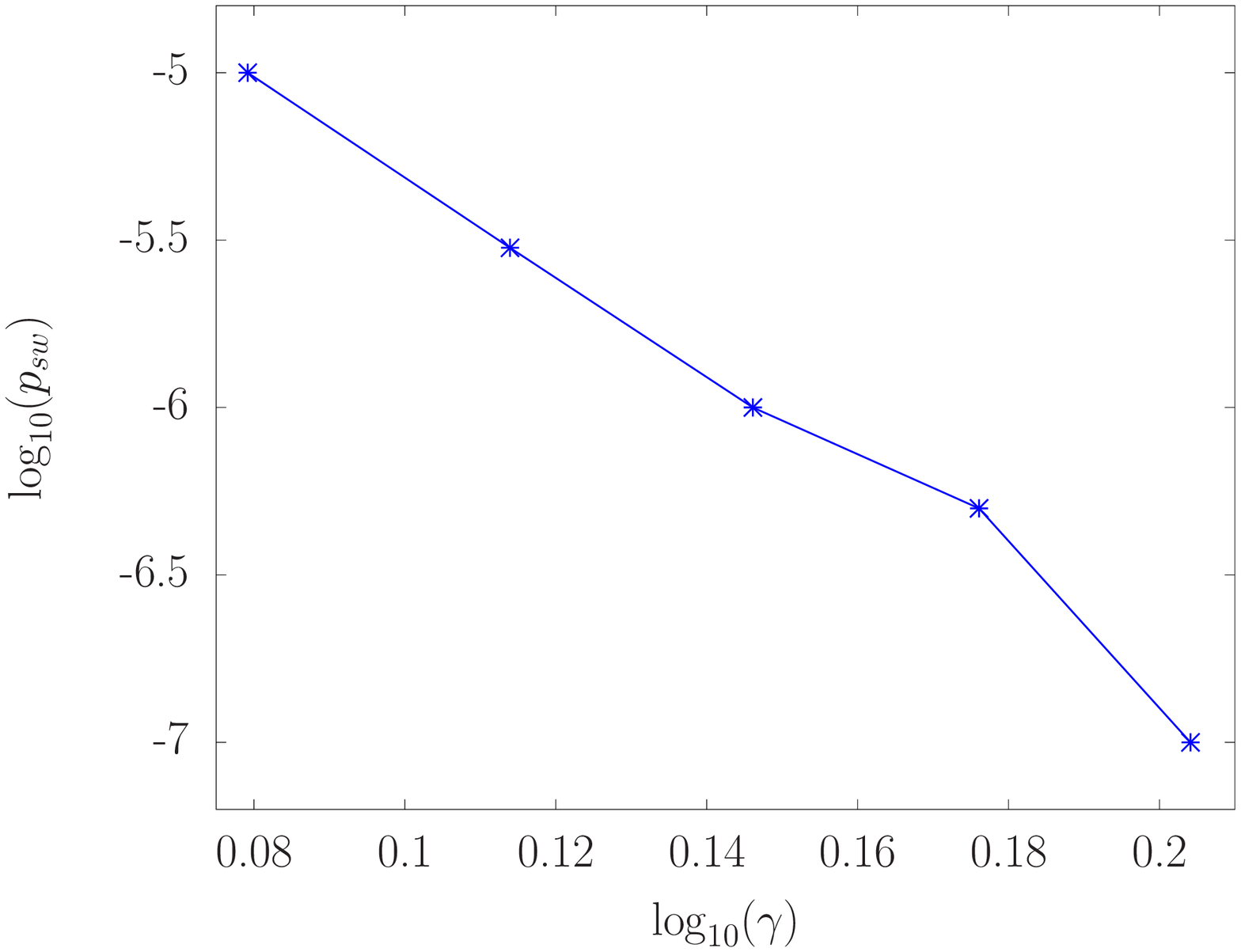}\\
(c)\includegraphics[width=8cm, keepaspectratio]{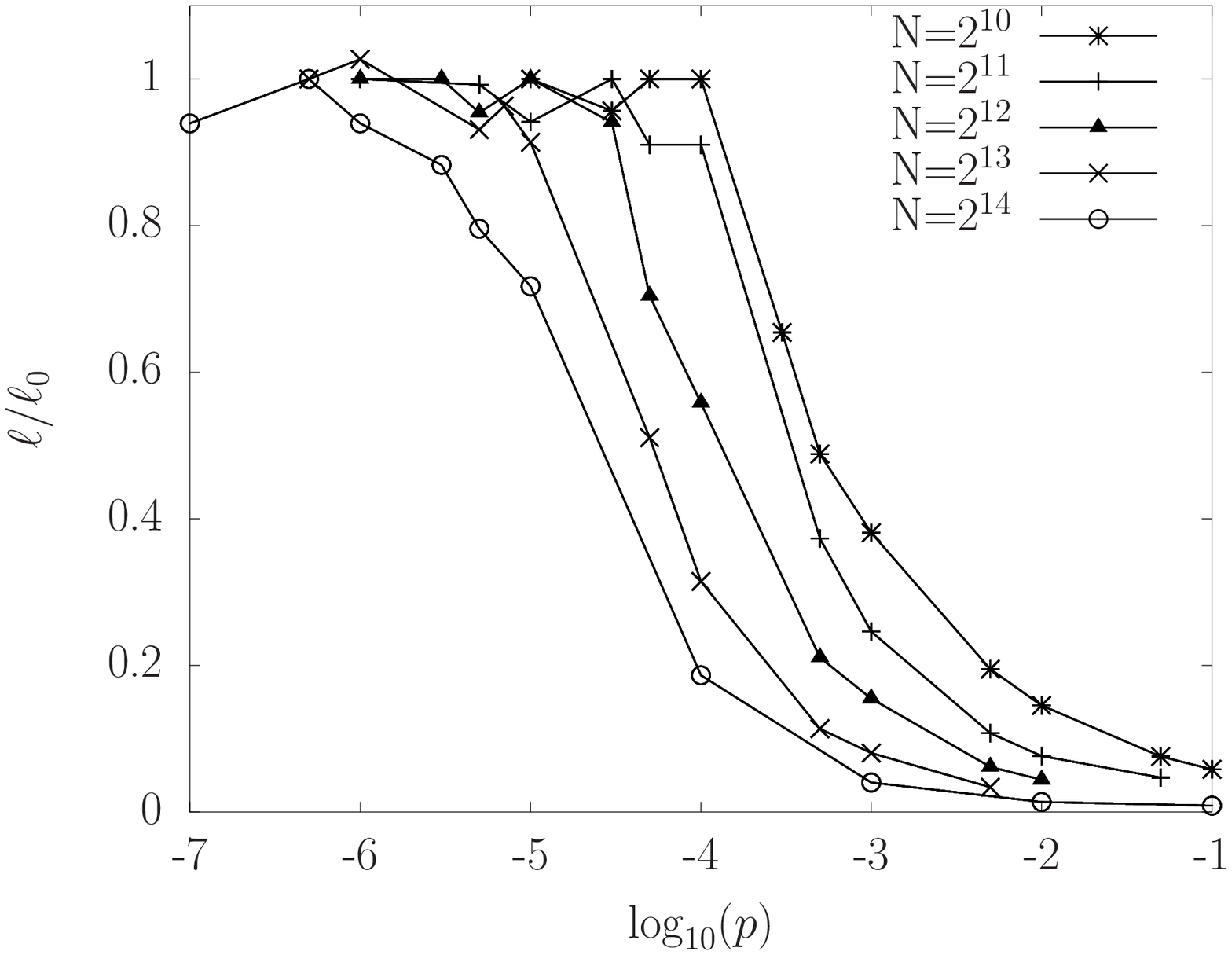}\\
(d)\includegraphics[width=8cm, keepaspectratio]{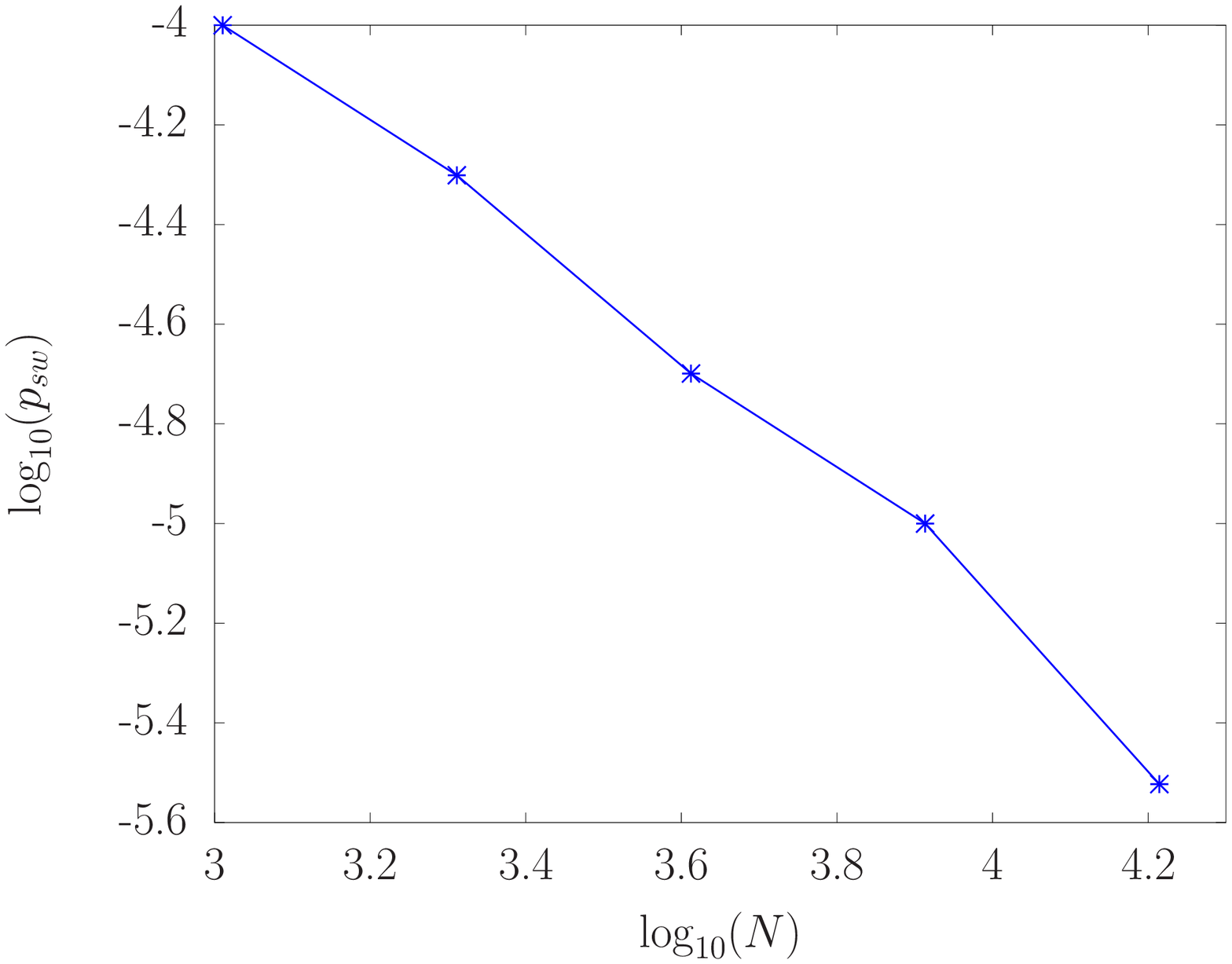}
\caption{(a) Average path lengths versus rewiring probability
for different $\gamma$ values and $N=2^{14}$. (b) Power law scaling of $p_{SW}(\gamma)$. 
(c) Average path lengths versus rewiring probability for different $N$ values and $\gamma=1.3$ . 
(d) Power law scaling of $p_{SW}(\gamma=1.3)$. The curve slope is $\approx 1.27$, 
coherent with the scaling in Eq.~(\ref{eq: pSW scaling}).
\label{fig:length}}
\end{figure}

The small-world regime embeds characteristics of both the regular
lattice and the random network ones: the network keeps track of the
initial configuration since, after the rewiring, it still conserves
a local neighborhood like a regular lattice; on the other hand the
network approaches, in the sense specified in Sec.~\ref{sec:The-Model},
the random graph topology because of the shortcuts induced by the
rewiring. In our context it hence emerges naturally the question of
how the degree, which scales as $k\sim N^{\gamma-1}$, could influence
the scaling of topological quantities in competition with the rewiring
probability $p$. For instance, a crucial passage in which the $\gamma$
parameter could play an important role is the crossover from the regular
chain topology to the small-world regime. It is usually investigated
by the scaling behavior of the average path length $l(p,\gamma)$,
defined as the average shortest distance between spins. 
\begin{figure}
\includegraphics[scale=0.4, keepaspectratio]{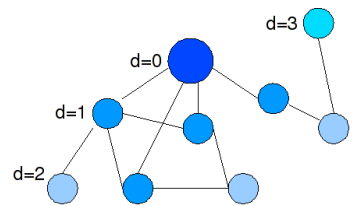}\caption{(color online) Path lengths starting from the blue vertex.\label{fig: path}}
\end{figure}
This quantity has an algebraic increase $l\sim N$ for a regular one-dimensional
lattice with fixed degree $k$, while for random networks it grows
as $l\sim\log N$. The passage between those two regimes is enhanced
by the long-range connections which could allow the spins to behave
coherently. Practically, since the network lacks a metric, the distance
between two spins is calculated as the minimal number of edges to
cross to go from one spin to the other, as shown in Fig.~\ref{fig: path}.
To investigate the change between these two behaviors we perform numerical
simulations, varying $\gamma$ and $p$: we use values for $\gamma$
from 1.2 to 1.5 and $p$ ranges from $10^{-7}$ up to $10^{-3}$ .
$N$ is fixed at $2^{14}$ and we average over 10 network realizations
for each value of $p$. In Fig.~\ref{fig:length}a we plot $l(\gamma,p)/l(\gamma,0)$
versus $\gamma$. $l(\gamma,p)$ shows the known crossover behavior
\cite{watts_strogatz1998SW} but, considering the probability $p_{SW}(N,p,\gamma)$
at which $l(\gamma,p)$ drops abruptly to the random network values,
it appears evident that it is strongly dependent on $\gamma$. We
have the following scaling for $p_{SW}(N,\gamma)$ \cite{newman1999scaling},
using the degree definition in Eq.~(\ref{eq:degree}):
\begin{equation}
p_{SW}\sim\frac{1}{N^{D}kD}\propto\left(\frac{1}{N}\right)^{\gamma},\label{eq: pSW scaling}
\end{equation}
where $D=1$ is the dimension of the initial regular lattice. In Fig.
\ref{fig:length}b we plot the estimation of $p_{SW}(N,\gamma)$ from
the simulations versus $\gamma$ which effectively confirm the power
law of Eq.~(\ref{eq: pSW scaling}). The degree is hence crucial
to quantitatively determine the passage to the small-world regime;
this dependence unveils its importance considering that, on small-world
networks, a ''topological'' length scale can be defined \cite{newman1999scaling}
as 
\begin{equation}
\xi=1/(pkD)^{1/D}\label{eq:correlation}
\end{equation}
and then $p_{SW}$ in Eq.~(\ref{eq: pSW scaling}) is the probability
of having $\xi=N$ . This is the key condition to achieve global coherence
and it clearly appears that the density of links, governed by the
parameter $\gamma$, and the randomness injected by $p$ concur in
complexifying the network topology. In Sec. \ref{sec:Thermodynamic-Behaviour regular},
we then move one step further dealing with the thermodynamics of the $XY$-rotors
model on the small-world network and looking for the topological signature
of the $\gamma$ and $p$ parameters in its properties.

\subsection{Thermodynamic Behavior on Small World Networks\label{sec:Thermodynamic-Behaviour}}

In Sec.~\ref{sub:Network-analysis} we focused on the topological
interplay of $\gamma$ and $p$ parameters in establishing the small-world
regime which, as explained, is noteworthy for its ambivalence, resembling
both to a regular lattice and to a random graph. In this section we
put the $XY$-rotors model on a small-world network: the question
we address now is to investigate the thermodynamic counterpart of
the network complex topology. We focus the low $\gamma$ regime, i.e.
$\gamma<1.5$. In this case we recall that the degree is still too
low to induce long-range order by itself without the intervention
of randomness and the network behaves like a one-dimensional chain.
In the interval $\gamma>1.5$ the high degree already induces a mean
field phase transition of the magnetization whose critical energy
is $\varepsilon_{c}=0.75$, as shown in Sec.~\ref{sec:Thermodynamic-Behaviour regular}.
In this interval, even without the contribution of long-range connections,
the network is connected enough to behave like a full-coupled one,
which is the case of the Hamiltonian Mean Field model. On the other
hand, in the case of random networks, it has been shown that the mean
field phase transition appears for all $\gamma>1$ \cite{ciani2011long}.
We thus introduce progressively long-range connections with the rewiring
probability $p$ since, from Eq.~(\ref{eq:correlation}), we expect
to retrieve two regimes determined by $\gamma$ and $p$; $\xi>N$
in which long-range order is absent and $\xi<N$ where the order parameter
displays a second order phase transition. 
\begin{figure}
(a)\includegraphics[width=8cm, keepaspectratio]{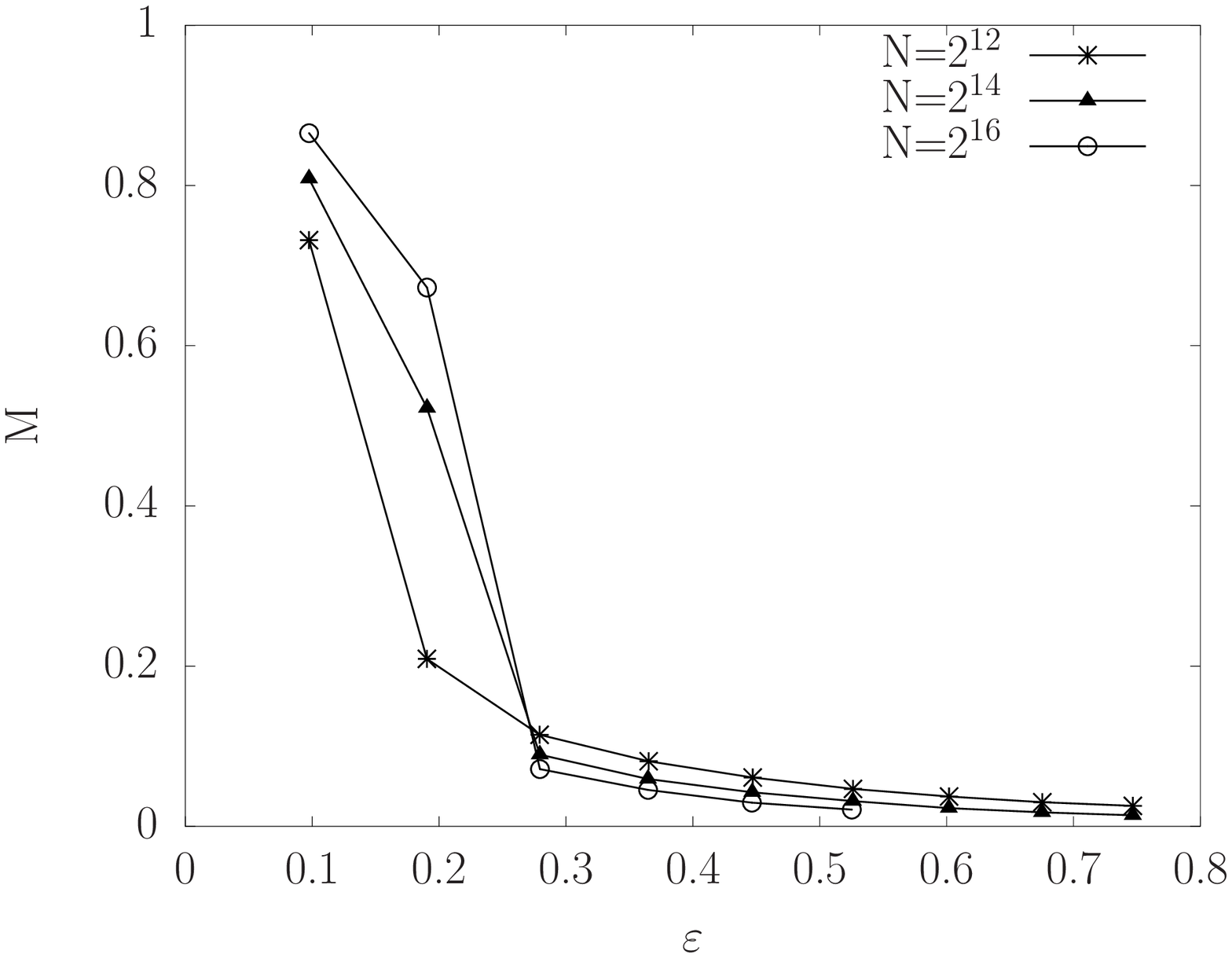}

(b)\includegraphics[width=8cm, keepaspectratio]{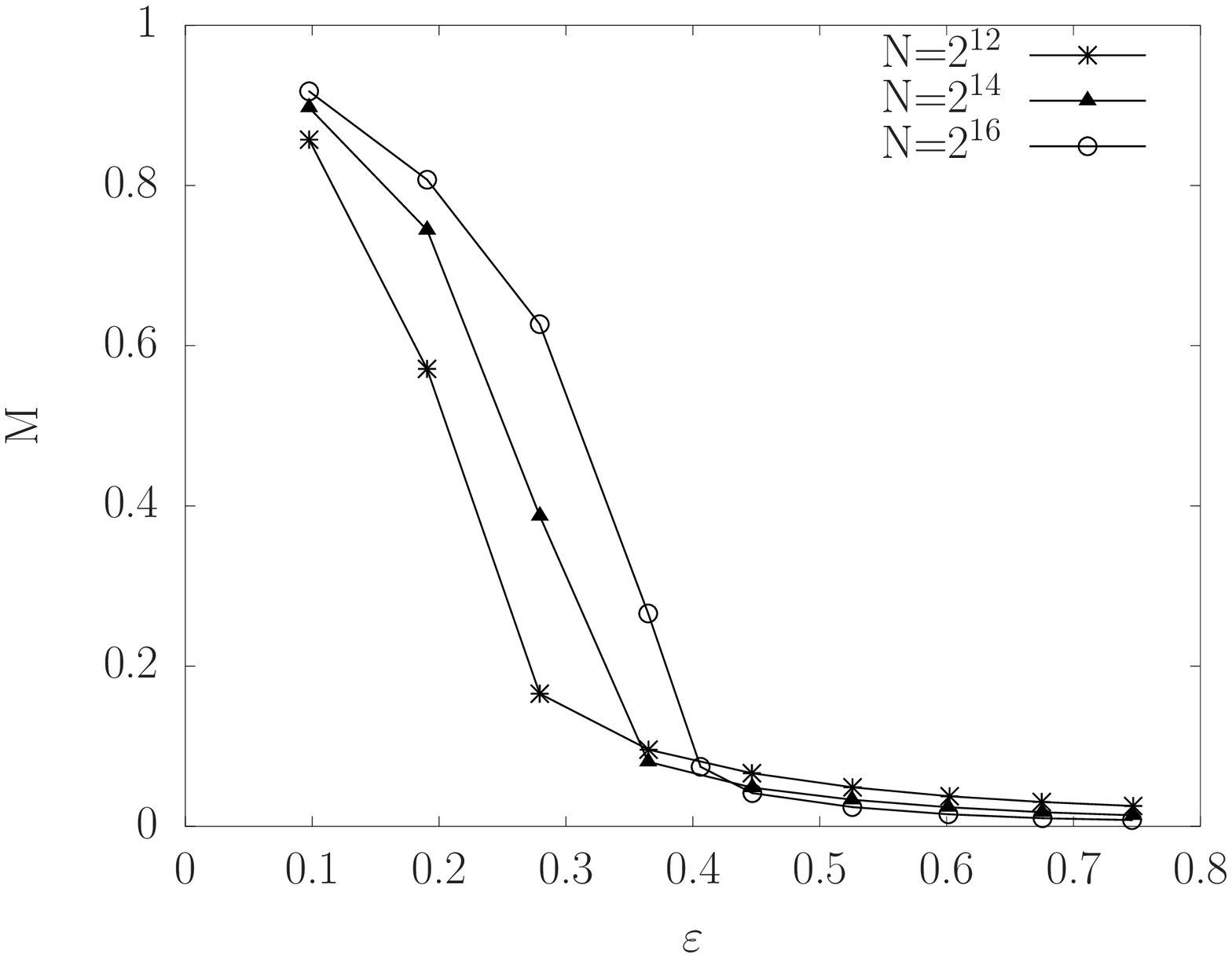}\\

(c)\includegraphics[width=8cm,, keepaspectratio]{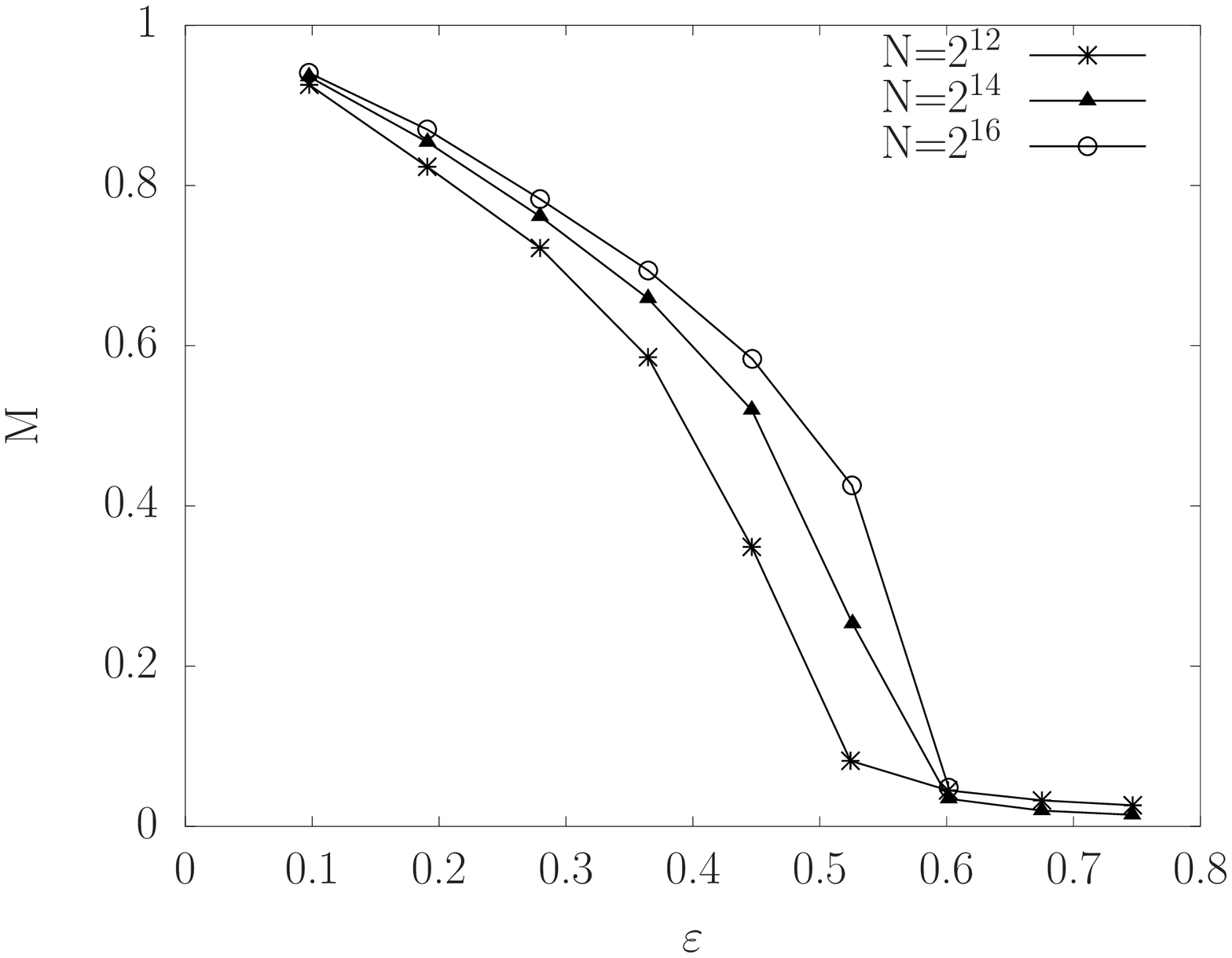}

\caption{Average magnetization $M$ versus energy density $\varepsilon=E/N$
for several system sizes, $\gamma=1.25$ and $p=0.001\,\,(a),\,0.005\,\,(b),\,0.05\,\,(c).$
\label{fig:different probability}}
\end{figure}
 In Figs.~\ref{fig:different probability}a-c we set $\gamma=1.25$
and, for each value of $p$, we consider several system sizes, above
and below the threshold $\xi(1.25,p,N)=N$. The results displayed
in Figs.~\ref{fig:different probability} show the equilibrium mean
value of the magnetization $\overline{M}$ versus the energy density
$\varepsilon=E/N$: for $N=2^{12}$, the probabilities $p=0.001$
and $0.005$ (Figs.~\ref{fig:different probability} a-b) are still
too low to entail the crossover to the long-range regime and the system
does not undergo a phase transition. On the other hand, the other
two sizes considered $N=2^{14}$ and $2^{16}$ are in the $\xi<N$
regime and the mean field phase transition is recovered all the $p$
taken in account. As explained, increasing the randomness decreases
the small-world threshold; hence all the sizes show the phase transition
of the magnetization for $p=0.05$ (Fig. \ref{fig:different probability}c).
Those results suggest the importance of $\xi$ also from the statistical point of view: in Sec.~\ref{sub:Network-analysis},
we showed that it signals the topological passage from regular to
small-world network which identifies itself by a drop of the average
path distance $l(N,\gamma,p)$; equivalently in this short $l(N,\gamma,p)$
regime the existence of long-range order is possible and, thus, we
observe the second field phase transition of the thermodynamic order
parameter. Remarkably, the critical energy $\varepsilon_{c}$ at which
the transition occurs varies accordingly to the randomness; we thus
investigate this effect tuning $\gamma$ between $1.2$ and $1.5$
and $p$ from $10^{-7}$ to $10^{-3}$. As explained before, it is
worth focusing on the interval $\gamma\leq1.5$. In this case the
shortcuts introduced by the rewiring process are crucial for the achievement
of global coherence; while in the $\gamma>1.5$ we already know that
phase transition with $\varepsilon_{c}=\varepsilon_{HMF}=0.75$ occurs
both on regular chains \cite{deNigris2013} and on random networks
\cite{ciani2011long}.
\begin{figure}
(a)\includegraphics[width=8cm, keepaspectratio]{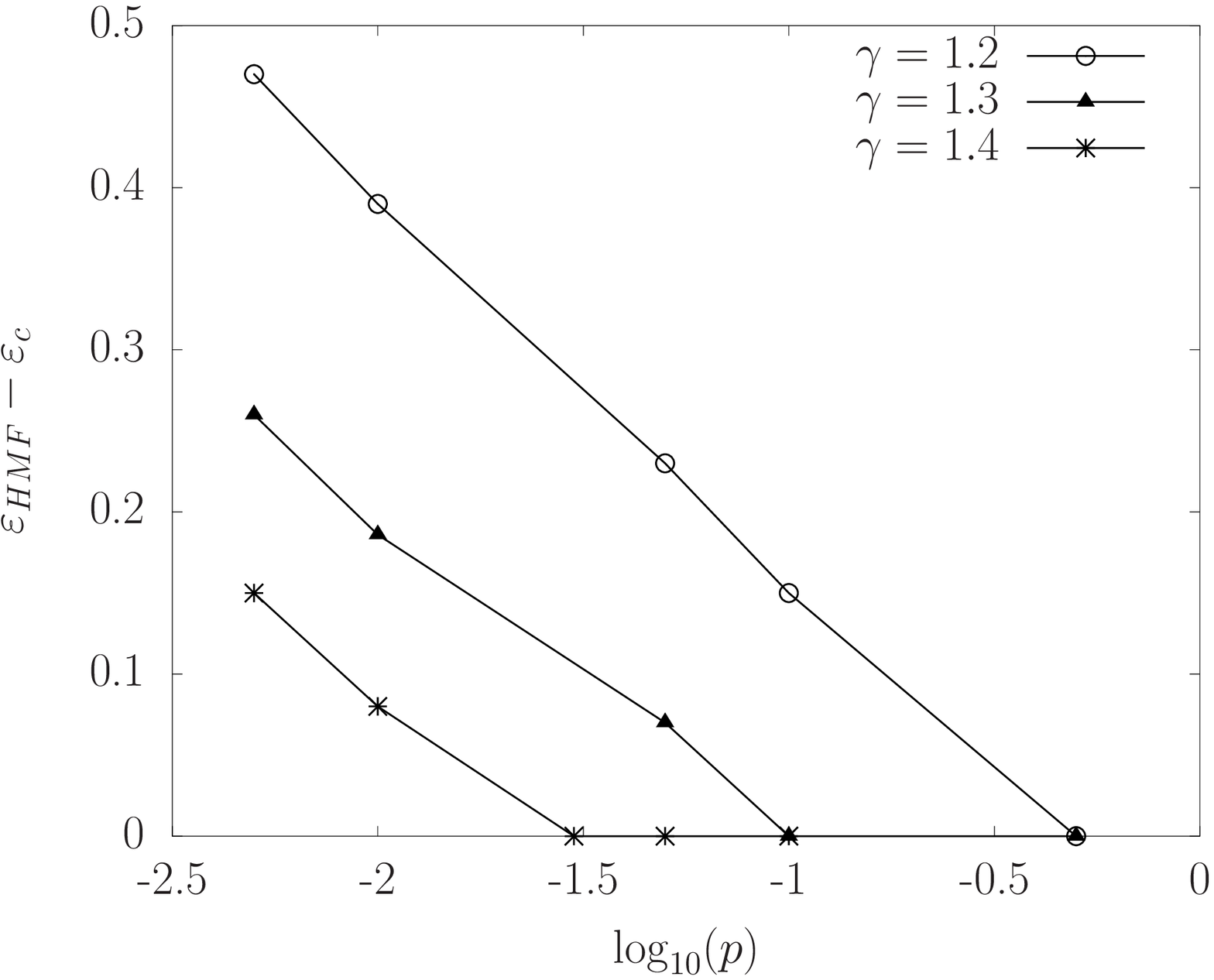}\\
(b)\includegraphics[width=8cm, keepaspectratio]{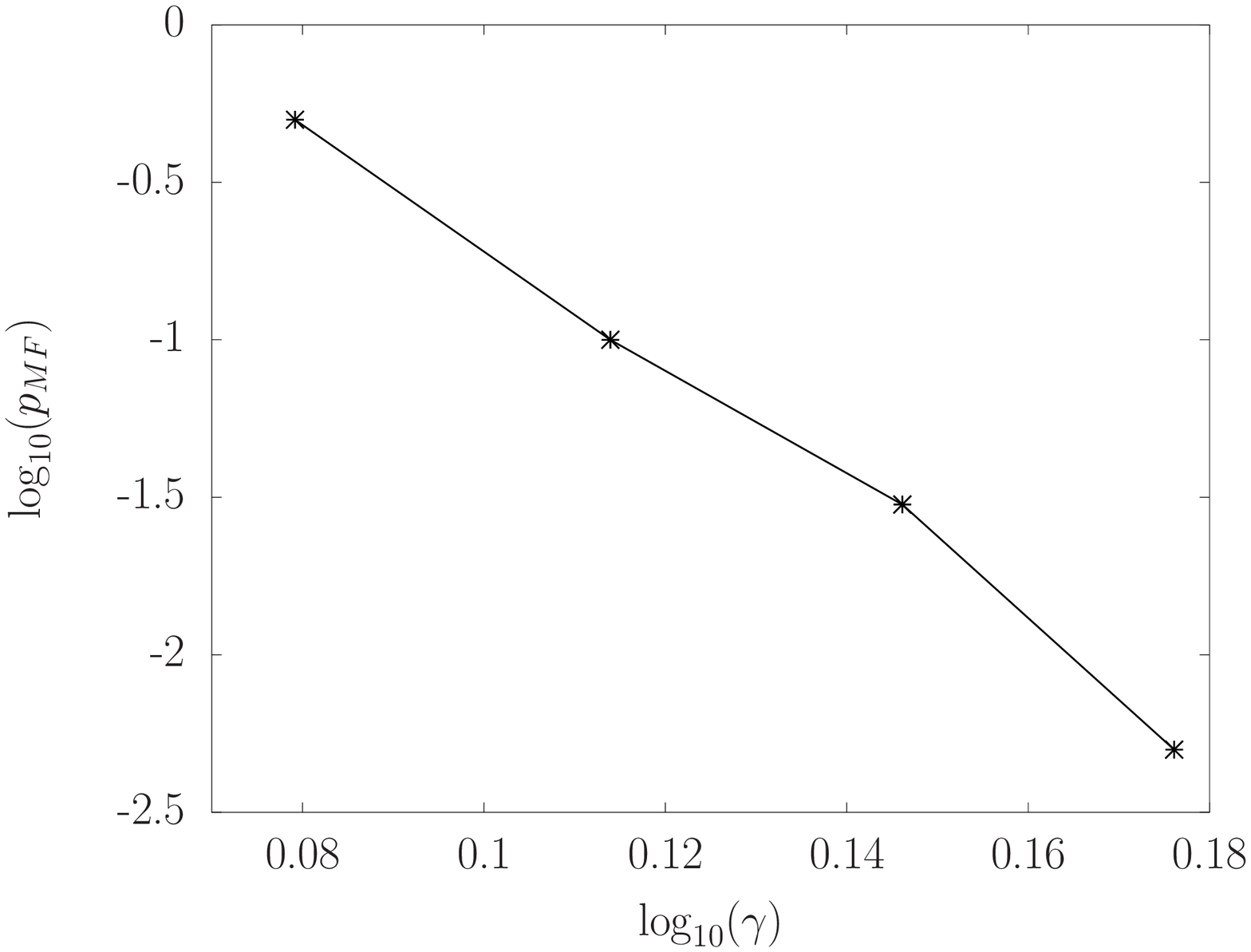}\\
(c)\includegraphics[width=8cm, keepaspectratio]{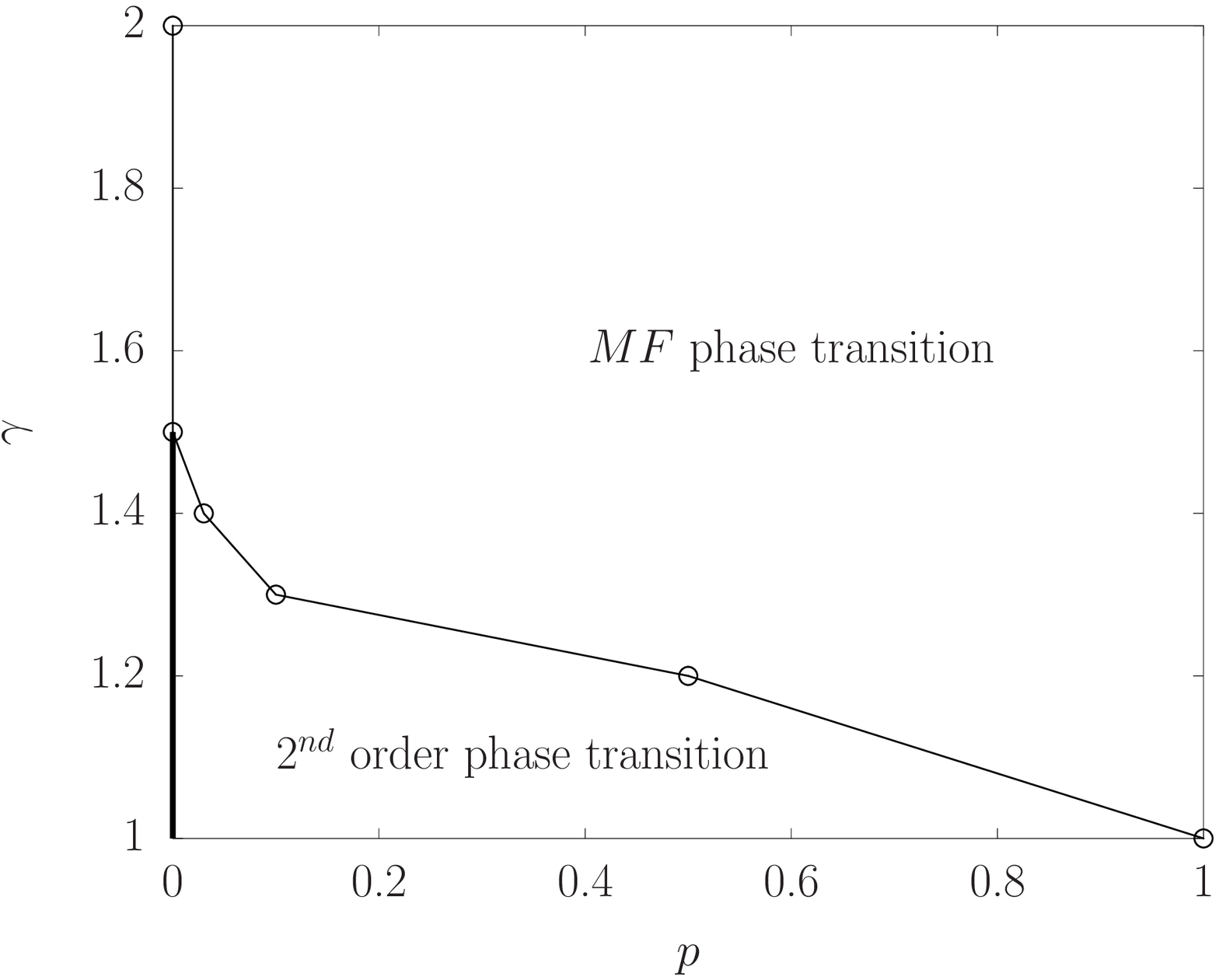}
\caption{(a) Logarithmic dependence of the critical energy $\varepsilon_{c}$
versus the rewiring probability $p$ for different $\gamma$ values. (b) Power law scaling of $p_{MF}(\gamma)$. (c) Phase plot in the $\left (\gamma,p\right )$ plane. The thick line for $p=0$ and $1<\gamma$<1.5 
stands for the absence of phase transitions in that parameter region. In the ``$MF$ phase transition`` region, the critical energy is the same of $HMF$, $\varepsilon_{c}=0.75$. \label{fig:log dependence}}
\end{figure}
 In Fig. \ref{fig:log dependence}, we plot the critical energy $\varepsilon_{c}(p,\gamma)$
versus the rewiring probability $p$ for several values of $\gamma$
and we observe that the phase boundary seems to be well described
by the logarithmic form :
\begin{equation}
\varepsilon_{c}=\log(g(\gamma)p^{c})\label{eq:critical energy}
\end{equation}
with $C\sim0.1$. Eq.~(\ref{eq:critical energy}) is coherent with
the scaling proposed in \cite{kim_smallworld2001,medvedyeva2003dynamic}
as far as the $p$ dependence is concerned. Remarkably, in \cite{kim_smallworld2001,medvedyeva2003dynamic},
it was a result issued from Monte-Carlo simulations in the canonical
ensemble while we work in the microcanonical frame. Moreover the aforementioned
results of logarithmic scaling were found in the $p\rightarrow0$ regime,
while where we are exploring regions with large values of $p$. We
also have to insist on the fact that Eq.~(\ref{eq:critical energy})
embeds an extra piece of information concerning the degree. Indeed,
in our analysis the ``quantitative'' topological parameter $\gamma$
affects in his turn the critical energy $\varepsilon_{c}$ through
the function $g(\gamma)$ , showing the non trivial role played by
the links density in the thermodynamic behavior of the $XY$-rotors
model. 

Another specific information can be retrieved from Fig.~\ref{fig:log dependence}a.
There is a threshold beyond which a ``saturation'' process exists:
to be more explicit, for each value of $\gamma$, we define a threshold
probability $p_{MF}(\gamma)$ for which the critical energy is $\varepsilon_{c}=0.75$,
identical to values obtained in the mean field ($\gamma=2$), or for
the fully randomized networks \cite{ciani2011long}. For $p>p_{MF}(\gamma)$
increasing the randomness does not influence anymore the critical
energy and in some way the resulting small-world network is, from
the thermodynamic point of view, equivalent to a fully coupled graph.
In Fig.~\ref{fig:log dependence}b, we show how this probability
threshold $p_{MF}(\gamma)$ depends as a power law on the $\gamma$
parameter. Note though that we expect that $p_{MF}\rightarrow0$ when
$\gamma\rightarrow1.5$, because as we discussed before in the $\gamma>1.5$
regime the system is already in the mean field state without any rewiring.
Regarding Fig.~\ref{fig:log dependence}b, we do not expect the results
to be valid near $\gamma=1.5$. Indeed a precise estimation of $p_{MF}$
proves to be very delicate since it relies in its turn on the determination
of the critical energy of the transition which is intrinsically a
hard task. Moreover the simulations are performed with finite size systems, so the measured $p_{MF}$ is influenced by finite size
effects. Then we actually have $p_{MF}(\gamma,N)$ and this dependence on $N$ can entail a finite value even for $\gamma=1.5$.
We also have to mention that we average on a finite number
of network realizations, which may affect as well the results. An interesting path to follow in order to avoid this effects and refine our estimations could be the use of finite-size scaling
techniques. Moreover, since previous results exist in the canonical ensemble \cite{medvedyeva2003dynamic,kim_smallworld2001}, this analysis would be of interest in our approach
because we deal with the microcanonical ensemble. It would then be possible to compare the characteristics of the phase transition in the two ensembles and shed light on
their equivalence. Proceeding
in our analysis, we recall that for the regular network a metastable
state was found for $\gamma_{c}\simeq 1.5$ in which the order parameter
is affected by heavy fluctuations, suggesting that the system oscillates
between low magnetization values, proper of the $\gamma<1.5$ regime,
and the mean field value of the $\gamma>1.5$ case. We can notice
that, after the introduction of randomness, we do not observe this
metastable state for $\gamma=1.5$ or any other value of $\gamma$.
In fact now a small (eventually vanishing) $p$ is enough to generate
a phase transition. It therefore exists an interplay between the ``quantitative''
parameter $\gamma$ and the ``qualitative'' parameter $p$; nevertheless
those parameters, as anticipated in Sec.~\ref{sec:The-Model}, are
not equivalent when dealing with their influence on the thermodynamic
behavior of the $XY$-model. This duality is so far not complete since
it was not possible to retrieve the metastable state in the $\gamma<1.5$
regime acting exclusively on the $p$ parameter. In this sense, the
randomness is ``regularizing'' the thermodynamic behavior: the rewired
network supports either the behavior of a regular lattice either,
once the small-world regime is reached, gives rise to the phase transition
of the magnetization. Summarizing, we can say that the noise created
by the rewiring stabilizes the passage between the two regimes and
destroys the delicate metastable state which arose in the regular
lattice.

\section{Conclusion}

In conclusion, we have studied the influence on the critical behavior
of the $XY$-rotors model of two different network topologies, the regular
lattice and small world network. In Sec.~\ref{sec:Thermodynamic-Behaviour regular},
we introduced the parameter $\gamma$ which allows to tune the number
of links from the linear chain to the full coupling configuration.
We identified two main parameter regions: the first for $\gamma<1.5$
in which the model has a one dimensional behavior and thus it does
not display long-range or quasi long-range order as shown by numerical
simulations. On the contrary, in the second region $(\gamma>1.5)$,
the spin degree is sufficiently high to lead to the emergence of a
coherent state: we thus observe a mean field phase transition of the
magnetization, identical to the one of the HMF model. More interestingly,
we show numerical and analytical evidence of an unstable state at
the threshold between the two regions, for $\gamma_{c}\simeq1.5$. In this
peculiar state, the magnetization is affected by fluctuations which
seem to be size independent and, furthermore, this state does not
reach equilibrium on the timescales considered. We then calculated
analytically an approximated expression for the magnetization, obtained
in the low temperatures regime, which demonstrates the topological
critical nature of $\gamma_{c}\simeq1.5$. This expression retrieves correctly
the two behaviors aforementioned and, since it contains the spectrum
of the adjacency matrix, it points out that the\emph{ topological
}origin of the three different phases shown by the simulations. We
have then studied the role of the links density on the topology of
small-world networks and its effect on the $XY$-rotors model dynamics.
We have focused, in Sec.~\ref{sec:The-Model}, on the crossover to
the small-world regime tuning the $\gamma$ parameter. We show by
numerical simulations that $p_{SW}$ has the scaling in Eq.~(\ref{eq: pSW scaling})
which is therefore consistent with \cite{newman1999scaling}. Hence
the links density, governed by $\gamma$, turns out to be crucial
to enhance the crossover between the ``large-world'' regime and
the small-world one cooperating with the rewiring probability $p$
in the creation of long-range connections. We then investigated, in
Sec.~\ref{sec:Thermodynamic-Behaviour}, the thermodynamic response
of the $XY$-rotors model to the variations of the network underlying.
We retrieved the emergence of a mean field transition of the magnetization
once $p>p_{SW}$. This latter condition implies the network to be
in the $\xi<N$ case, using the definition in Eq.~(\ref{eq:correlation}), implying that the 
passage between the regular and the small-world topology also entails a difference in the behaviour of the model. Moreover we found
a logarithmic dependence of the critical energy $\varepsilon_{c}(p,\gamma)$
on $p$ and $\gamma$ which lead to the scaling in Eq.~(\ref{fig:log dependence}).
The interplay between the topological parameters in modifying $\varepsilon_{c}$
saturates when $\varepsilon_{c}=0.75$ which is the critical energy
of the Hamiltonian Field model and we defined a new threshold
probability $p_{MF}$ which displays the power law scaling with $\gamma$
shown in Fig.~\ref{fig:log dependence}b. We hence found that a small
(vanishing) amount of randomness regularizes the $\gamma=1.5$ metastable
state pointed out in Sec.~\ref{sec:Thermodynamic-Behaviour regular}
and moreover it was not possible to recreate it in the $\gamma<1.5$
interval just adding long-range connections with $p$. Therefore,
as far as the thermodynamic behavior is concerned, we conclude that
$\gamma$ and $p$ are not equivalent when dealing with the transition
to the mean field state; nevertheless, we anticipate here that a more
refined criteria than randomness could be found in order to perturb
the regular network in the low density regime $(\gamma<1.5)$ and
enhance the creation of out-of-equilibrium effects like the $\gamma_{c}\simeq1.5$
metastable state.

\begin{acknowledgments}
X. L. is partially supported by the FET project Multiplex 317532 and
S. d. N. is supported by DGA/MRIS.
\end{acknowledgments}

\bibliographystyle{apsrev4-1}
%\bibliography{networks}

%

\end{document}